\documentclass[letterpaper, 10pt]{IEEEtran}%
\usepackage{cite}
\usepackage{amsmath,amssymb,amsfonts}
\usepackage{algorithmic}
\usepackage{graphicx}
\usepackage{textcomp}
\usepackage{xcolor}
\usepackage{mathrsfs}
\usepackage{psfrag}
\usepackage{cancel}
\usepackage{cuted}
\def\mc{\mathcal}

\def\BibTeX{{\rm B\kern-.05em{\sc i\kern-.025em b}\kern-.08em
    T\kern-.1667em\lower.7ex\hbox{E}\kern-.125emX}}
\begin{document}
\title{{\LARGE \bf Necessary and Sufficient Conditions for State Feedback Equivalence to Negative Imaginary Systems} \\
}

\author{Kanghong Shi,$\quad$Ian R. Petersen, \IEEEmembership{Fellow, IEEE},$\quad$and$\quad$Igor G. Vladimirov 
\thanks{This work was supported by the Australian Research Council under grant DP190102158.}
\thanks{K. Shi, I. R. Petersen and I. G. Vladimirov are with the School of Engineering, College of Engineering and Computer Science, Australian National University, Canberra, Acton, ACT 2601, Australia.
        {\tt kanghong.shi@anu.edu.au}, {\tt ian.petersen@anu.edu.au}, {\tt igor.vladimirov@anu.edu.au}.}%
}

\newtheorem{definition}{Definition}
\newtheorem{theorem}{Theorem}
\newtheorem{conjecture}{Conjecture}
\newtheorem{lemma}{Lemma}
\newtheorem{remark}{Remark}
\newtheorem{corollary}{Corollary}
\newtheorem{assumption}{Assumption}

\maketitle

\thispagestyle{empty}
\pagestyle{empty}

\begin{abstract}
In this paper, we present necessary and sufficient conditions under which a linear time-invariant (LTI) system is state feedback equivalent to a negative imaginary (NI) system. More precisely, we show that a minimal LTI strictly proper system can be rendered NI using full state feedback if and only if it can be output transformed into a system, which has relative degree less than or equal to two and is weakly minimum phase. We also considered the problems of state feedback equivalence to output strictly negative imaginary systems and strongly strict negative imaginary systems. Then we apply the NI state feedback equivalence result to robustly stabilize an uncertain system with strictly negative imaginary uncertainty. An example is provided to illustrate the proposed results, for the purpose of stabilizing an uncertain system.
\end{abstract}


\section{INTRODUCTION}
Negative imaginary (NI) systems theory was introduced in \cite{lanzon2008stability,petersen2010feedback} and has attracted attention in the past decade \cite{xiong2010negative,song2012negative,mabrok2014generalizing,wang2015robust,bhikkaji2011negative,bhowmick2017lti}. Motivated by the control of flexible structures \cite{preumont2018vibration,halim2001spatial,pota2002resonant}, NI systems theory has been applied in many fields including nano-positioning control \cite{mabrok2013spectral,das2014mimo,das2014resonant,das2015multivariable} and the control of lightly damped structures \cite{cai2010stability,rahman2015design,bhikkaji2011negative}, etc. Typical mechanical NI systems are systems with colocated force actuators and position sensors. In this sense, NI systems theory provides an alternative to positive real (PR) systems theory \cite{brogliato2007dissipative}, as PR systems theory uses negative velocity feedback control while NI systems theory uses positive position feedback control. In comparison with PR systems theory, one advantage of NI systems theory is that it allows systems to have relative degrees of zero, one and two, while PR systems can only have relative degrees of zero and one.

Roughly speaking, a square transfer matrix is NI if it is stable and its Hermitian imaginary part is negative semidefinite for all frequencies $\omega \geq 0$. For a single-input single-output (SISO) NI system, its frequency response has a phase lag between $0$ to $2\pi$ radians for all frequencies $\omega>0$. It is shown using a set of linear matrix inequalities (LMIs) in the NI lemma that a system is NI if it is dissipative, with the supply rate being the inner product of its input and the derivative of its output \cite{xiong2010negative,song2012negative,mabrok2020dissipativity}. An NI system $R(s)$ can be robustly stabilized using a positive feedback strictly negative imaginary (SNI) controller $R_s(s)$, where $R(\infty)R_s(\infty)=0$ and $R_s(\infty)\geq 0$, if and only if the DC loop gain of the interconnection is strictly less than unity; i.e., $\lambda_{max}(R(0)R_s(0))<1$; see \cite{lanzon2008stability}.

The problem of rendering a system PR using state feedback control in order to achieve stabilization has been investigated in many papers (see \cite{kokotovic1989positive,saberi1990global}, etc). For example, \cite{saberi1990global} renders a linear system PR and this result is then generalized to nonlinear systems in \cite{byrnes1991passivity} using passivity theory. Further nonlinear generalizations of these ideas are presented in the papers \cite{byrnes1991asymptotic,santosuosso1997passivity,lin1995feedback,jiang1996passification}. In these papers, such PR or passivity state feedback equivalence results are then applied to stabilize systems with specific nonlinearities. One of the necessary and sufficient conditions for state feedback equivalence to a passive or PR system is that the original system must have relative degree one. This restriction stems from the nature of passivity and PR systems and, as a result, rules out a wide variety of control systems with relative degree two, such as mechanical systems with force actuators and position sensors. To overcome this limitation and to complement the existing results that are based on passivity and PR systems theory, we consider the problem of state feedback equivalence to NI systems.

In this paper, we investigate the conditions under which a linear system with the minimal realization $(\mc A, \mc B, \mc C)$ is state feedback equivalent to an NI system. Suppose the system has no zeros at the origin. We show that such a system can be rendered NI via the use of state feedback if and only if (a) it can be output transformed into a system with relative degree less than or equal to two; and (b) the transformed system is weakly minimum phase (see for example \cite{isidori2013nonlinear} for details of the terminology in feedback stabilization). The idea of applying an output transformation comes from the fact that the system in question does not always have a relative degree vector in general and hence does not always have a normal form. However, we show that the property of NI state feedback equivalence is invariant to a nonsingular output transformation because its effect can be compensated by an additional input transformation. Moreover, we show that a system can be rendered output strictly negative imaginary (OSNI) if and only if it can be rendered NI. In particular, we show that a system is state feedback equivalent to a strongly strict negative imaginary (SSNI) system if and only if it has a relative degree vector $\{1,\cdots,1\}$ and is minimum phase. The proposed NI state feedback equivalence results are then applied to robustly stabilize an uncertain system with SNI uncertainty.

The contribution of this paper is to provide conditions under which a system is state feedback equivalent to an NI system, an OSNI system or an SSNI system. This work, together with the preliminary conference paper \cite{shi2021negative}, is the first in the literature where NI state feedback equivalence is investigated. In \cite{shi2021negative}, we consider cases where a system has relative degree of either one or two, which rules out the case that a system has mixed relative degrees one and two. In this paper, we consider the general case which allows the system to have mixed relative degrees. Also, the relative degree condition is an assumption in \cite{shi2021negative}, while it is a part of the necessary and sufficient conditions in this paper. This makes the present paper a complete result for the NI state feedback equivalence problem. This paper also contributes to the literature by providing a method to stabilize systems with relative degree less than or equal to two.

This paper is organised as follows: Section \ref{section:pre} provides the essential background on NI systems theory. Section \ref{section:NI state feedback equivalence} contains the main results of this paper, where we derive necessary and sufficient conditions under which it is possible to render a system NI using state feedback control. Formulas for the required state feedback matrices are provided in the proofs. In Section \ref{section:SSNI}, an SSNI state feedback equivalence result is also provided. Section \ref{section:synthesis} applies the NI state feedback equivalence results presented in Section \ref{section:NI state feedback equivalence} in stabilizing an uncertain system with SNI uncertainty. Section \ref{section:example} illustrates the presented results with a numerical example. Section \ref{section:conclusion} concludes the paper.
 
\textbf{Notation}: The notation in this paper is standard. $\mathbb R$ and $\mathbb C$ denote the fields of real and complex numbers, respectively. $\mathbb N$ denotes the set of nonnegative integers. $j\mathbb R$ denotes the set of purely imaginary numbers. $\mathbb R^{m\times n}$ and $\mathbb C^{m\times n}$ denote the spaces of real and complex matrices of dimension $m\times n$, respectively. $\Re [\cdot]$ is the real part of a complex number. $A^T$ and $A^*$ denote the transpose and complex conjugate transpose of a matrix $A$, respectively. $A^{-T}$ denotes the transpose of the inverse of $A$; i.e., $A^{-T}=(A^{-1})^T=(A^T)^{-1}$. $ker(A)$ denotes the kernel of A. $spec(A)$ denotes the spectrum of $A$. $\lambda_{max}(A)$ denotes the largest eigenvalue of a matrix $A$ with real spectrum. For a symmetric or Hermitian matrix $P$, $P>0\ (P\geq 0)$ denotes the property that the matrix $P$ is positive definite (positive semidefinite) and $P<0\ (P\leq 0)$ denotes the property that the matrix $P$ is negative definite (negative semidefinite). For a positive definite matrix $P$, we denote by $P^{\frac{1}{2}}$, the unique positive definite square root of $P$. $OLHP$ and $CLHP$ are the open and closed left half-planes of the complex plane, respectively.

\section{PRELIMINARIES}\label{section:pre}

\begin{definition}(Negative Imaginary Systems)\label{def:NI}
\cite{xiong2010negative}
A square real-rational proper transfer function matrix $R(s)$ is said to be negative imaginary if:

1. $R(s)$ has no poles at the origin and in $\Re [s]>0$;

2. $j[R(j\omega)-R^*(j\omega)]\geq 0$ for all $\omega \in (0,\infty)$ except for values of $\omega$ where $j\omega$ is a pole of $R(s)$;

3. if $j\omega_0$ with $\omega_0\in (0,\infty)$ is a pole of $R(s)$, then it is a simple pole and the residue matrix $K_0=\lim_{s\to j\omega_0}(s-j\omega_0)jR(s)$ is Hermitian and positive semidefinite.
\end{definition}

\begin{definition}(Strictly Negative Imaginary Systems)\cite{xiong2010negative}
A square real-rational proper transfer function matrix $R(s)$ is said to be strictly negative imaginary if the following conditions are satisfied:

1. $R(s)$ has no poles in $\Re[s]\geq 0$;

2. $j[R(j\omega)-R^*(j\omega)]> 0$ for all $\omega \in (0,\infty)$.	
\end{definition}

\begin{definition}(Output Strictly Negative Imaginary Systems)\label{def:OSNI}\cite{bhowmick2019output}
A square real-rational proper transfer function matrix $R(s)$ is said to be output strictly negative imaginary if there exists a scalar $\epsilon > 0$ such that
\begin{equation*}
j\omega [R(j\omega)-R(j\omega)^*]-\epsilon \omega^2 \bar R(j\omega)^*\bar R(j\omega) \geq 0
\end{equation*}
$\forall \omega \in \mathbb R \cup {\infty}$ where $\bar R(j\omega) = R(j\omega)-R(\infty)$. In this case, we say $R(s)$ is OSNI with a level of output strictness $\epsilon$.
\end{definition}

\begin{definition}(Strongly Strictly Negative Imaginary Systems)\label{def:SSNI}
\cite{lanzon2011strongly}
A square real-rational proper transfer function matrix $R(s)$ is said to be strongly strictly negative imaginary if the following conditions are satisfied:

1. $R(s)$ is SNI.

2. $\lim_{\omega \to \infty}j\omega[R(j\omega)-R^*(j\omega)]>0$ and $\lim_{\omega \to 0}j\frac{1}{\omega}[R(j\omega)-R^*(j\omega)]>0$.
\end{definition}

\begin{lemma}(NI Lemma)\cite{xiong2010negative}\label{lemma:NI}
Let $(A,B,C,D)$ be a minimal state-space realisation of an $p\times p$ real-rational proper transfer function matrix $R(s)$ where $A\in \mathbb R^{n\times n}$, $B\in \mathbb R^{n\times p}$, $C\in \mathbb R^{p\times n}$, $D\in \mathbb R^{p\times p}$. Then $R(s)$ is NI if and only if:

1. $\det(A)\neq 0$, $D=D^T$;

2. There exists a matrix $Y=Y^T>0$, $Y\in \mathbb R^{n\times n}$ such that
\begin{equation*}
	AY+YA^T\leq 0,\qquad \textnormal{and} \qquad B+AYC^T=0.
\end{equation*}	
\end{lemma}

\begin{lemma}(SSNI Lemma)\cite{lanzon2011strongly}\label{thm:SSNI}
Given a square transfer function matrix $R(s)\in \mathbb R^{p\times p}$ with a state-space realisation $(A,B,C,D)$, where $A\in \mathbb R^{n\times n}$, $B\in \mathbb R^{n\times p}$, $C \in \mathbb R^{p\times n}$ and $D\in \mathbb R^{p\times p}$. Suppose $R(s)+R(-s)^T$ has normal rank $p$ and $(A,B,C,D)$ has no observable uncontrollable modes. Then $A$ is Hurwitz and $R(s)$ is SSNI if and only if $D=D^T$ and there exists a matrix $Y=Y^T>0$ that satisfies conditions
\begin{equation*}
AY+YA^T<0, \quad \textnormal{and} \quad B+AYC^T=0.
\end{equation*}
\end{lemma}

\begin{lemma}(OSNI Lemma)\cite{bhowmick2019output}\label{lem:OSNI}
Let $(A,B,C,D)$ be a minimal state-space realisation of an $p\times p$ real-rational proper transfer function matrix $R(s)$ where $A\in \mathbb R^{n\times n}$, $B\in \mathbb R^{n\times p}$, $C\in \mathbb R^{p\times n}$, $D\in \mathbb R^{p\times p}$. Let $\epsilon>0$ be a scalar. Then $R(s)$ is OSNI with a level of output strictness $\epsilon$ if and only if $D=D^T$ and there exists a matrix $Y=Y^T>0$, $Y\in \mathbb R^{n\times n}$ such that
\begin{equation*}
	AY+YA^T+\epsilon(CAY)^TCAY \leq 0,\quad \textnormal{and} \quad B+AYC^T=0.
\end{equation*}

\end{lemma}

\begin{definition}(Lyapunov Stability)\cite{bernstein2009matrix}
\label{def:LS}
A square matrix $A$ is said to be Lyapunov stable if $spec(A)\subset CLHP$ and every purely imaginary eigenvalue of $A$ is semisimple.
\end{definition}

\begin{lemma}(Lyapunov Stability Theorem - Asymptotic Stablity)\cite{hespanha2018linear}\label{thm:Lyapunov}
Consider a continuous-time homogeneous linear time-invariant (LTI) system
\begin{equation}\label{eq:LTI system}
\dot x = \mc A x,\qquad x \in \mathbb R^n	,
\end{equation}
the following statements are equivalent:

1. The system (\ref{eq:LTI system}) is asymptotically stable.

2. All of the eigenvalues of $\mc A$ have strictly negative real parts.

3. For every symmetric positive definite matrix $\mc Q$, there exists a unique solution $\mc P$ to the following Lyapunov equation
\begin{equation}\label{eq:Lyapunov}
	\mc A^T\mc P+\mc P\mc A = -\mc Q
\end{equation}
such that $\mc P$ is symmetric and positive definite.

4. There exists a symmetric positive definite matrix $\mc P$ for which the following Lyapunov matrix inequality holds:
\begin{equation*}
	\mc A^T\mc P+\mc P\mc A<0.
\end{equation*}
\end{lemma}

\begin{lemma}(Lyapunov Stability Theorem - Lyapunov Stablity)\cite{bernstein2009matrix}\label{thm:marginally stable}
	Let $\mc A\in \mathbb R^{n\times n}$ and assume there exists a positive semidefinite matrix $\mc Q \in \mathbb R^{n\times n}$ and a positive definite matrix $\mc P \in \mathbb R^{n\times n}$ such that (\ref{eq:Lyapunov}) is satisfied, then $\mc A$ is Lyapunov stable.
\end{lemma}

\begin{lemma}(Eigenvector Test for Controllability)\cite{hespanha2018linear}\label{thm:eigenvector test for ctrl}
The pair $(A,B)$ is controllable if and only if there is no eigenvector of $A^T$ in the kernel of $B^T$.
\end{lemma}

\begin{lemma}(Eigenvector Test for Observability)\cite{hespanha2018linear}\label{thm:eigenvector test for obsv}
The pair $(A,C)$ is observable if and only if no eigenvector of $A$ is in the kernel of $C$.
\end{lemma}

\begin{lemma}(Internal Stability of Interconnected NI Systems)\cite{xiong2010negative}\label{lemma:dc gain theorem}
	Consider an NI transfer function matrix $R(s)$ and an SNI transfer function matrix $R_s(s)$ that satisfy $R(\infty)R_s(\infty)=0$ and $R_s(\infty)\geq 0$. Then the positive feedback interconnection $[R(s),R_s(s)]$ is internally stable if and only if $\lambda_{max}(R(0)R_s(0))<1$. (e.g., see \cite{lanzon2008stability} for the definition of internal stability and positive feedback interconnection.)
\end{lemma}

\section{STATE FEEDBACK EQUIVALENCE TO AN NI SYSTEM}
\label{section:NI state feedback equivalence}

Consider a system with the state-space model:
\begin{subequations}\label{eq:original}
\begin{align}
\dot x =& \ \mc A x + \mc  B u,\\
y =& \ \mc C x, 
\end{align}
\end{subequations}
where $x \in \mathbb R^{n}$ is the state, $u \in \mathbb R^p$ is the input and $y \in \mathbb R^p$ is the output. Here, $\mc A \in \mathbb R^{n\times n}$, $\mc B \in \mathbb R^{n\times p}$ and $\mc C \in \mathbb R^{p\times n}$. We assume that $rank(\mc B) = rank (\mc C)=p$.

For the system (\ref{eq:original}), we provide the following definitions.

\begin{definition}(see also \cite{fomichev2016generalization, isidori2013nonlinear})\label{def:RD vector}
A vector $r=\{r_1,\cdots, r_p\}\in \mathbb N^p$ is called the relative degree vector of system (\ref{eq:original}) if the following conditions are satisfied.

1. For all $i=1,\cdots,p$,
\begin{align}
& \mc C_i \mc A^{j} \mc B=0 \quad \textnormal{for} \quad j=0,\cdots,r_i-2;\notag\\
& \textnormal{and} \quad H(r)_i:=\mc C_i\mc A^{r_i-1}\mc B\neq 0.\label{eq:rd vector condition 1}
\end{align}

2. $\det (H(r))\neq 0$.

Here $\mc C_i$ denotes the $i$-th row of the matrix $\mc C\in \mathbb R^{p\times n}$
and 
\begin{equation}\label{eq:H(r)}
H(r)=\left[\begin{matrix}
	\mc C_1 \mc A^{r_1-1}\mc B\\
	\vdots\\
	\mc C_p \mc A^{r_p-1}\mc B
	\end{matrix}\right].
\end{equation}
\end{definition}
Condition 1 in this definition indicates that the $i$-th output has its $r_{i}$-th time derivative depending explicitly on the inputs.

As is explained in \cite{fomichev2016generalization}, in the case that (\ref{eq:original}) is a MIMO system; i.e., $p\geq 2$, Condition 2 in Definition \ref{def:RD vector} is not always satisfied. The components in the relative degree vector $r$ are invariant under a nonsingular state transformation. However, a nonsingular output transformation can change the components in the vector $r$ and in some cases transform the realization $(\mc A,\mc B,\mc C)$ to $(\mc A, \mc B, \tilde {\mc C})$, where $\tilde {\mc C} = T_y \mc C$, $T_y \in \mathbb R^{p\times p}$ and $\det (T_y)\neq 0$, which satisfies Condition 2 in Definition \ref{def:RD vector}.

Note that there does not always exist such an output transformation that transforms the system (\ref{eq:original}) into a form with a relative degree vector. In \cite{fomichev2016generalization}, the notion of a leading incomplete relative degree vector is introduced as follows.

\begin{definition}\cite{fomichev2016generalization}\label{def:LIRD}
A vector $r=\{r_1,...,r_p\}\in \mathbb N^p$ is called a leading incomplete relative degree (LIRD) vector of the system (\ref{eq:original}) if the following conditions are satisfied.

1. For all $i=1,\cdots,p$,
\begin{equation*}
\begin{aligned}
& \mc C_i \mc A^{j} \mc B=0 \quad \textnormal{for} \quad j=0,\cdots,r_i-2;\\
& \textnormal{and} \quad \mc C_i\mc A^{r_i-1}\mc B\neq 0.
\end{aligned}
\end{equation*}

2. $r_i \leq r_{i+1}$ for $i=1,\cdots,p-1$.

3. For any set of pairwise distinct indices $i_1,\cdots, i_q \in \{1,2,\cdots,p\}$ such that $r_{i_1}=r_{i_2} = \cdots = r_{i_q}$, the rows $H(r)_{i_1},\cdots,H(r)_{i_q}$ are linearly independent, where $H(r)$ is defined in (\ref{eq:H(r)}) and $H(r)_i$ is defined in (\ref{eq:rd vector condition 1}).
\end{definition}

As is explained in \cite{fomichev2016generalization} and \cite{kraev2014generalization}, if a LIRD vector is such that all rows in $H(r)$ are linearly independent, then this LIRD vector is a relative degree vector as defined in Definition \ref{def:RD vector}. This relationship can also be observed by comparing Definitions \ref{def:RD vector} and \ref{def:LIRD}.

\begin{lemma}\label{lem:always reducible}\cite{kraev2014generalization}
For any controllable system with the realization $(\mc A, \mc B,\mc C)$, there exists a nonsingular output transformation such that the transformed system has an LIRD vector.  
\end{lemma}
\begin{IEEEproof}
This follows directly from Remark 4 and Lemma 4 in \cite{kraev2014generalization}.	
\end{IEEEproof}

In this paper, we derive conditions for the NI state feedback equivalence of the system (\ref{eq:original}) by investigating the normal form of an auxiliary system, which is obtained by applying an output transformation to the original system. This leads to a transformed system with a relative degree vector. We show later in this paper that the existence of such an output transformation is one of the necessary conditions for NI state feedback equivalence. First, let us provide the definition for state feedback equivalence to an NI system.

\begin{definition}\label{def:state feedback equivalent NI}
A system in the form of (\ref{eq:original}) is said to be state feedback equivalent to an NI system if there exists a state feedback control law
\begin{equation*}
u=K_x x+ K_v v,	
\end{equation*}
where $K_x\in \mathbb R^{p\times n}$ and $K_v \in \mathbb R^{p\times p}$, is such that the resulting closed-loop system with the new input $v\in \mathbb R^p$, is minimal and NI.
\end{definition}

Note that state feedback equivalence problems do not allow for a change of output. However, they allow for a free change of inputs. We show in the following two lemmas that for a system of the form of (\ref{eq:original}), its NI state feedback equivalence property is invariant to a nonsingular output transformation.

\begin{lemma}\label{lemma:NI property preservation}
Suppose $T\in \mathbb R^{p\times p}$ is nonsingular. Then the transfer matrix $T R(s) T^T$ is NI if and only if $R(s)$ is NI.
\end{lemma}
\begin{IEEEproof}
The proof is based on Definition \ref{def:NI}. $R(s)$ is NI if and only if Conditions 1, 2 and 3 in Definition \ref{def:NI} are satisfied. However, the positive definiteness (semi-definiteness) of the matrices in Conditions 1, 2 and 3 in Definition \ref{def:NI} is invariant to the transformation $T R(s) T^T$. This completes the proof.
\end{IEEEproof}

\begin{lemma}\label{lem:feedback NI irrelevance}
Consider the system (\ref{eq:original}) and the state, input and output transformations $\tilde x = T_x x$, $\tilde u = T_u u$ and $\tilde y = T_y y$, where $T_x \in \mathbb R^{n\times n}$, $T_u \in \mathbb R^{p\times p}$ and $T_y \in \mathbb R^{p\times p}$ are nonsingular. Then the system (\ref{eq:original}) is state feedback equivalent to an NI system if and only if the transformed system is also state feedback equivalent to an NI system.
\end{lemma}
\begin{IEEEproof}
If the transformed system with state $\tilde x$, input $\tilde u$ and output $\tilde y$ is state feedback equivalent to an NI system, then there exists a control law
\begin{equation*}
\tilde u = K_x \tilde x + K_v \tilde v,	
\end{equation*}
under which the system with input $\tilde v$ and output $\tilde y$ is minimal and NI. According to Lemma \ref{lemma:NI property preservation}, now the system with output $ y = T_y^{-1}\tilde y$ and input $ v = T_y^T \tilde v$ is also minimal and NI. This means that the original system with state $ x$, input $ u$ and output $ y$ is also state feedback equivalent to an NI system. The corresponding feedback control law can be derived as shown in the following:
\begin{equation*}
 u = T_u^{-1}\tilde u = T_u^{-1} (K_x \tilde x +K_v \tilde v)=T_u^{-1} (K_x T_x x +K_v T_y^{-T} v).
\end{equation*}
This completes the sufficiency part of the proof. Since the state, input and output transformation matrices $T_x$, $T_u$ and $T_y$ are all nonsingular, the necessity part of the proof follows in the same manner as the sufficiency part with the inverses of the transformations considered.
\end{IEEEproof}

In this paper, we consider systems whose relative degree vector only consists of numbers less than or equal to two, as we show later that this is one of the necessary conditions for state feedback equivalence to NI systems.

\begin{definition}
The system (\ref{eq:original}) is said to have relative degree less than or equal to two if it has a relative degree vector $r=\{r_1,\cdots,r_p\}$, where $1 \leq r_i\leq 2$ for all $i=1,\cdots,p$.
\end{definition}

Consider the case that there exists an output transformation $\tilde y=T_y  y$, where $T_y \in \mathbb R^{p\times p}$ and $\det(T_y)\neq 0$, that transforms the system (\ref{eq:original}) into a form with a relative degree vector $r=\{r_1,\cdots,r_p\}$. Let $\tilde {\mc C} = T_y \mc C \in \mathbb R^{p\times n}$, then the transformed system takes the form:
\begin{subequations}\label{eq:output transformed system}
\begin{align}
\dot {x} =&\ \mc A  x + \mc B  u,\\
\tilde y =&\ \tilde {\mc C} x,		
\end{align}
\end{subequations}
where $rank(\mc B)=rank(\tilde {\mc C})=p$. 

\begin{lemma}\label{lemma:normal form}
Suppose the system (\ref{eq:output transformed system}) has relative degree less than or equal to two. Then there exist input and state transformations that transform (\ref{eq:output transformed system}) into the following normal form:
\begin{subequations}\label{eq:original normal}
\begin{align}
\dot z =&\ A_{00}z+A_{01}x_1+A_{02}x_2+A_{03}x_3,\label{eq:internal}\\
\dot x_1 =& A_{10}z + A_{11}x_1+A_{12}x_2+A_{13}x_3+u_1,\label{eq:rd1}\\
\dot x_2 =& x_3,\label{eq:rd2}\\
\dot x_3 =& A_{30}z+A_{31}x_1+A_{32}x_2+A_{33}x_3 + u_2,\label{eq:rd2d}\\
\tilde y=&\left[\begin{matrix}
x_1 \\ x_2	
\end{matrix}\right].
\end{align}
\end{subequations}
where $\tilde x=\left[\begin{matrix}z \\ 
x_1 \\ x_2 \\ x_3
\end{matrix}\right]\in \mathbb R^{n}$ is the state, $\tilde u=\left[\begin{matrix}u_1 \\ u_2 \end{matrix}\right]\in \mathbb R^{p}$ is the input and $\tilde y\in \mathbb R^p$ is the output of the transformed system. Here, $x_1, u_1 \in \mathbb R^{p_1}$ and $x_2,x_3, u_2 \in \mathbb R^{p_2}$, where $0\leq p_1\leq p$ and $p_2:=p-p_1$. Also, $z\in \mathbb R^{m}$, where $m:=n-p-p_2$.
\end{lemma}
\begin{IEEEproof}
Without loss of generality, suppose the components in the relative degree vector $r$ of the system (\ref{eq:output transformed system}) are sorted in nondecreasing order, i.e., $r=\{1,\cdots,1, 2\cdots,2\}$. Let $p_1$ $(0\leq p_1 \leq p)$ be the number of ones in $r$ and $p_2=p-p_1$ be the number of twos in $r$. Also, define the following matrices:
\begin{equation*}
\tilde {\mc C}_O = \left[\begin{matrix}
	\tilde {\mc C}_1\\
	\vdots\\
	\tilde {\mc C}_{p_1}
	\end{matrix}\right]\in \mathbb R^{p_1 \times n},\quad \textnormal{and} \quad \tilde {\mc C}_T = \left[\begin{matrix}
	\tilde {\mc C}_{p_1+1}\\
	\vdots\\
	\tilde {\mc C}_{p}
	\end{matrix}\right]\in \mathbb R^{p_2 \times n},
\end{equation*}
where $\tilde {\mc C}_i$ is the $i$-th row in the matrix $\tilde {\mc C}$. Hence, $\tilde {\mc C}_O$ is the block matrix in $\tilde {\mc C}$ which determines the output entries corresponding to the ones in $r$. $\tilde {\mc C}_T$ is the block matrix in $\tilde {\mc C}$ which determines the output entries corresponding to the twos in $r$. According to Definition \ref{def:RD vector}, we have that $rank(\tilde {\mc C}_O \mc B) = p_1$, $\tilde {\mc C}_T \mc B =0$ and $rank(\tilde {\mc C}_T \mc A \mc B) = p_2$. Also, Condition 2 in Definition \ref{def:RD vector} implies that
\begin{equation}\label{eq:independent rd property}
	\det\left[\begin{matrix}
	\tilde {\mc C}_O\mc B\\
	\tilde {\mc C}_T\mc A\mc B
	\end{matrix}\right] \neq 0.
\end{equation}
Therefore, the rows of the matrix $\left[\begin{matrix}
	\tilde {\mc C}_O\\
	\tilde {\mc C}_T\mc A
	\end{matrix}\right]$ are linearly independent. Since $rank(\mc C)=p$ and $\det T_y\neq 0$, then $rank(\tilde {\mc C})=p$. Hence, $\tilde {\mc C}_T$ has full row rank. Also, according to Condition 1 in Definition \ref{def:RD vector}, we have that $\tilde {\mc C}_T\mc B = 0$. Then we can prove by contradiction that the rows of $\tilde {\mc C}_T$ are linearly independent of the rows of $\left[\begin{matrix}
	\tilde {\mc C}_O\\
	\tilde {\mc C}_T\mc A
	\end{matrix}\right]$. Indeed, suppose there exists a row $(\tilde {\mc C}_T)_\kappa$ of $\tilde C_T$, which is a linear combination of the rows of $\left[\begin{matrix}
	\tilde {\mc C}_O\\
	\tilde {\mc C}_T\mc A
	\end{matrix}\right]$. Then $(\tilde {\mc C}_T)_\kappa \mc B \neq 0$ according to (\ref{eq:independent rd property}), which contradicts the equation $\tilde {\mc C}_T \mc B=0$. Therefore, the matrix $\left[\begin{matrix}
	\tilde {\mc C}_O\\ \tilde {\mc C}_T \\
	\tilde {\mc C}_T\mc A \end{matrix}\right]$ has full row rank.	Define the new state as
	\begin{align*}
	x_1 =&\ \tilde {\mc C}_O  x,\\
	x_2 =&\ \tilde {\mc C}_T  x,\\
	x_3 =&\ \dot x_2 = \tilde {\mc C}_T \mc A  x.
	\end{align*}
We also need a complementary state $z \in \mathbb R^{m}$ where $m:=n-p-p_2 \geq 0$. Let $z = \tilde {\mc C}_z x$, where $\tilde {\mc C}_z$ is such that
\begin{equation*}
	T_x = \left[\begin{matrix}
	\tilde {\mc C}_z\\
	\tilde {\mc C}_O\\ \tilde {\mc C}_T \\
	\tilde {\mc C}_T\mc A \end{matrix}\right]
\end{equation*}
is nonsingular, and also $\tilde {\mc C}_z \mc B = 0$. Let $\tilde x = T_x x$ be the new state. Also, let
\begin{equation*}
\tilde u = \left[\begin{matrix}
	u_1 \\ u_2 \end{matrix}\right]=\left[\begin{matrix}
	\tilde {\mc C}_O \\ \tilde {\mc C}_T \mc A \end{matrix}\right]\mc B  u.
\end{equation*}
According to $(\ref{eq:independent rd property})$, the input transformation matrix $T_u = \left[\begin{matrix}
	\tilde {\mc C}_O \\ \tilde {\mc C}_T \mc A \end{matrix}\right]\mc B$ is nonsingular. The new system has a state-space model
\begin{subequations}\label{eq:original matrix form}
	\begin{align}
	\frac{d}{dt} \left[\begin{matrix}
	z \\ x_1 \\ x_2 \\x_3 \end{matrix}\right] =& T_x \mc A T_x^{-1} \left[\begin{matrix}
	z \\ x_1 \\ x_2 \\x_3 \end{matrix}\right] + \left[\begin{matrix}
	0 \\ u_1 \\ 0 \\u_2 \end{matrix}\right],\\
	\tilde y=& \left[\begin{matrix}
	0 & I & 0 & 0\\ 0& 0&I&0 \end{matrix}\right]\left[\begin{matrix}
	z \\ x_1 \\ x_2 \\x_3 \end{matrix}\right].
	\end{align}
\end{subequations}
By considering the blocks of $T_x \mc A T_x^{-1}$ including the relation $\dot x_2 = x_3$, we can write (\ref{eq:original matrix form}) in the form (\ref{eq:original normal}). This completes the proof.
\end{IEEEproof}

We now consider necessary and sufficient conditions under which the system (\ref{eq:original normal}) is state feedback equivalent to an NI system. For the system (\ref{eq:original normal}), choose the control inputs $u_1$ and $u_2$ to be
\begin{align}
u_1=&\ v_1+(K_{10}-A_{10})z+(K_{11}-A_{11})x_1\notag\\
&+(K_{12}-A_{12})x_2+(K_{13}-A_{13})x_3,\label{eq:u1}
\end{align}
and
\begin{align}
u_2=&\ v_2+(K_{20}-A_{30})z+(K_{21}-A_{31})x_1\notag\\
&+(K_{22}-A_{32})x_2+(K_{23}-A_{33})x_3,\label{eq:u2}
\end{align}
which allows the system (\ref{eq:original normal}) to be represented in the form
\begin{subequations}\label{eq:linearised2}
\begin{align}
\dot {\tilde x} =&\ A\tilde x+B\tilde v,\\
\tilde y=&\ C \tilde x,
\end{align}
\end{subequations}
where $\tilde v=\left[\begin{matrix}v_1 \\ v_2\end{matrix}\right]$ is the new input and
\begin{align}
A=&	\left[\begin{matrix}A_{00}&A_{01}&A_{02}&A_{03}\\K_{10}&K_{11} & K_{12} & K_{13} \\ 0&0&0&I \\ K_{20} & K_{21} & K_{22} & K_{23}\end{matrix}\right],\label{eq:rd1 A}\\
B=&\left[\begin{matrix}0 & 0\\ I & 0 \\ 0&0\\ 0& I\end{matrix}\right],\label{eq:rd1 B}\\
C=& \left[\begin{matrix}0& I&0&0 \\ 0&0&I&0\end{matrix}\right].\label{eq:rd1 C}
\end{align}

We need to find the state feedback matrices
\begin{align}
& K_{10}\in \mathbb R^{p_1\times m}, K_{11}\in \mathbb R^{p_1\times p_1}, K_{12}\in \mathbb R^{p_1\times p_2},\notag\\
& K_{13}\in \mathbb R^{p_1\times p_2},
 K_{20} \in \mathbb R^{p_2\times m}, K_{21} \in \mathbb R^{p_2\times p_1},\notag\\
& K_{22} \in \mathbb R^{p_2\times p_2},\ \textnormal{and}\  K_{23} \in \mathbb R^{p_2\times p_2}\label{eq:state feedback matrices}
\end{align}
such that the system (\ref{eq:linearised2}) is minimal and NI. The following lemma provides necessary and sufficient conditions for such state feedback matrices to exist.

\begin{lemma}\label{lem:feedback NI}
Suppose the system (\ref{eq:original normal}) satisfies $\det A_{00} \neq 0$. Then it is state feedback equivalent to an NI system if and only if it is controllable and $A_{00}$ is Lyapunov stable.
\end{lemma}
\begin{IEEEproof}
The system (\ref{eq:original normal}) is state feedback equivalent to an NI system if and only if there exist state feedback matrices (\ref{eq:state feedback matrices}) such that the system (\ref{eq:linearised2}) is NI and the realization $(A,B,C)$ in (\ref{eq:rd1 A})-(\ref{eq:rd1 C}) is minimal.

First, we prove that the controllability of the system (\ref{eq:original normal}) is equivalent to the controllability of the system (\ref{eq:linearised2}). Define 
\begin{equation*}
\breve A=	\left[\begin{matrix}A_{00}&A_{01}&A_{02}&A_{03}\\A_{10}&A_{11} & A_{12} & A_{13} \\ 0&0&0&I \\ A_{30} & A_{31} & A_{32} & A_{33}\end{matrix}\right].
\end{equation*}
Then we need to prove that the controllability of $(\breve A,B)$ is equivalent to that of $(A,B)$. According to Lemma \ref{thm:eigenvector test for ctrl}, the controllability of $(\breve A,B)$ implies that any non-zero vector in the kernal of $B^T$ is not an eigenvector of $\breve A^T$. Considering the structure of $B$ in (\ref{eq:rd1 B}), a non-zero vector $\eta \in \ker (B^T)$ must take the form $\small\eta = \left[\begin{matrix}\eta_1 \\ 0 \\ \eta_3 \\ 0\end{matrix}\right]$, where $\eta_1 \neq 0$ or $\eta_3\neq 0$. Therefore, for any scalar $\lambda_c$, we have that $\breve A^T \eta \neq \lambda_c \eta$. Substituting for $\breve A$, we obtain
\begin{equation}\label{eq:controllability condition}
	\left[\begin{matrix}A_{00}^T\eta_1 \\ A_{01}^T\eta_1 \\ A_{02}^T\eta_1 \\ A_{03}^T\eta_1+\eta_3\end{matrix}\right]\neq \lambda_c \left[\begin{matrix}\eta_1 \\ 0\\ \eta_3 \\ 0\end{matrix}\right]
\end{equation}
for any scalar $\lambda_c$. This condition depends only on the matrices $A_{00}$, $A_{01}$, $A_{02}$ and $A_{03}$, which forms the common first block row of the matrices $\breve A$ and $A$. Hence, the controllability of $(\breve A,B)$ is equivalent to that of $(A,B)$.

\textbf{Sufficiency.}
According to Lemma \ref{thm:eigenvector test for ctrl}, (\ref{eq:controllability condition}) is satisfied if and only if for any eigenvector $\eta_1$ of $A_{00}^T$ with eigenvalue $\lambda_c$, $A_{01}^T\eta_1\neq 0$ or $\left[\begin{matrix}A_{02}^T\eta_1 \\ A_{03}^T\eta_1+\eta_3\end{matrix}\right]\neq \lambda_c \left[\begin{matrix} \eta_3 \\ 0\end{matrix}\right]$. The condition $A_{01}^T\eta_1\neq 0$ holds if and only if $(A_{00},A_{01})$ is controllable. The condition $\left[\begin{matrix}A_{02}^T\eta_1 \\ A_{03}^T\eta_1+\eta_3\end{matrix}\right]\neq \lambda_c \left[\begin{matrix} \eta_3 \\ 0\end{matrix}\right]$ holds if and only if for any $\eta_3=-A_{03}^T\eta_1$, we have that $A_{02}^T\eta_1 \neq \lambda_c \eta_3 = -\lambda_c A_{03}^T\eta_1 = -A_{03}^TA_{00}^T\eta_1$. That is $(A_{03}^TA_{00}^T+A_{02}^T)\eta_1 \neq 0$, which holds if and only if $(A_{00},A_{00}A_{03}+A_{02})$ is controllable. Therefore, we conclude that $(A,B)$ is controllable if and only if $(A_{00},A_{01})$ or $(A_{00},A_{00}A_{03}+A_{02})$ is controllable.

We now derive necessary and sufficient conditions under which $(A,C)$ is observable. Given the structure of $C$ in (\ref{eq:rd1 C}), any non-zero vector $\sigma \in \ker (C)$ must take the form $\small\sigma = \left[\begin{matrix} \sigma_1 \\ 0 \\ 0  \\ \sigma_4\end{matrix}\right]$, where $\sigma_1 \neq 0$ or $\sigma_4 \neq 0$ . According to Lemma \ref{thm:eigenvector test for obsv}, $(A,C)$ is observable if and only if $A\sigma \neq \lambda_o \sigma$ for any scalar $\lambda_o$. Substituting $A$ from (\ref{eq:rd1 A}), we obtain
\begin{equation}\label{eq:observability condition}
	\left[\begin{matrix} A_{00}\sigma_1+A_{03}\sigma_4 \\ K_{10}\sigma_1 +K_{13}\sigma_4\\ \sigma_4  \\ K_{20}\sigma_1+K_{23}\sigma_4\end{matrix}\right]\neq \lambda_o \left[\begin{matrix} \sigma_1 \\ 0 \\ 0\\ \sigma_4\end{matrix}\right].
\end{equation}
When $\sigma_4 \neq 0$, (\ref{eq:observability condition}) is always true. Now we consider the case that $\sigma_1 \neq 0$ and $\sigma_4 = 0$. In this case, (\ref{eq:observability condition}) becomes
\begin{equation*}
	\left[\begin{matrix} A_{00}\sigma_1 \\ K_{10}\sigma_1 \\ 0 \\ K_{20}\sigma_1\end{matrix}\right]\neq \lambda_o \left[\begin{matrix} \sigma_1 \\ 0 \\ 0\\ 0\end{matrix}\right],
\end{equation*}
which holds if and only if for any vector $\sigma_1$ that is an eigenvector of $A_{00}$, $K_{10}\sigma_1 \neq 0$ or $K_{20}\sigma_1 \neq 0$. Therefore, according to Lemma \ref{thm:eigenvector test for obsv}, we conclude that $(A,C)$ is observable if and only if $(A_{00},K_{10})$ or $(A_{00},K_{20})$ is observable.

The nonsingular matrix $A_{00}$ is Lyapunov stable (see Definition \ref{def:LS}) if and only if there exists a state transformation $A_{00} \mapsto SA_{00}S^{-1}$ which allows $A_{00}$ to be represented, without loss of generality, as $A_{00}=diag(A_{00}^a,A_{00}^b)$, where
\begin{equation}\label{eq:A_00a A_00b definition}
\begin{aligned}
	& spec( A_{00}^a)\subset j\mathbb R\backslash \{0\}, \quad spec( A_{00}^b)\subset OLHP,\\
	& \textnormal{and} \quad A_{00}^a+( A_{00}^a)^T=0.
\end{aligned}
\end{equation}
Here $A_{00}^a \in \mathbb R^{m_a\times m_a}$ and $A_{00}^b \in \mathbb R^{m_b\times m_b}$, where $0\leq m_a \leq m$ and $m_b:=m-m_a$. The conditions in (\ref{eq:A_00a A_00b definition}) are achievable according to the proof of Proposition 11.9.6 in \cite{bernstein2009matrix}. Decomposing $A_{01}$, $A_{02}$, $A_{03}$, $K_{10}$ and $K_{20}$ accordingly using the same state-space transformation, we can write (\ref{eq:linearised2}) as
\begin{subequations}\label{eq:linearised2 expand}
\begin{align}
\dot z_1 =&\ A_{00}^a z_1+ A_{01}^a x_1+A_{02}^a x_2+A_{03}^a x_3,\\
\dot z_2 =&\ A_{00}^b z_2+ A_{01}^b x_1+A_{02}^b x_2+A_{03}^b x_3,\\
\dot x_1 =&\ K_{10}^a z_1 +K_{10}^b z_2 + K_{11}x_1+K_{12}x_2+K_{13}x_3+v_1,\\
\dot x_2 =&\ x_3,\\
\dot x_3 =&\ K_{20}^a z_1 + K_{20}^b z_2 +K_{21}x_1+K_{22}x_2+K_{23}x_3 + v_2,\\
y=&\ C\left[\begin{matrix}, 
z_1\\z_2\\x_1 \\ x_2\\x_3	
\end{matrix}\right],\ C=\left[\begin{matrix}0&0& I&0&0 \\ 0&0&0&I&0\end{matrix}\right]\label{eq:new C}.
\end{align}
\end{subequations}
Since $A_{00}^b$ is Hurwitz, there exist $\mc Y_1^b=(\mc Y_1^b)^T>0$ and $Q_b=Q_b^T>0$ such that
\begin{equation*}
	A_{00}^b\mc Y_1^b+\mc Y_1^b (A_{00}^b)^T=-Q_b.
\end{equation*}
Let $K_{20}$ be defined as
\begin{equation}\label{eq:rd1 K_20}
	K_{20}= \left[\begin{matrix}K_{20}^a & K_{20}^b\end{matrix}\right],
\end{equation}
where
\begin{equation}\label{eq:K20a}
	K_{20}^a=-{A_{02}^a}^T(A_{00}^a)^{-T}-(A_{03}^a)^T,
\end{equation}
and
\begin{equation}\label{eq:K20b}
	K_{20}^b=\left(-{A_{02}^b}^T(A_{00}^b)^{-T}-(A_{03}^b)^T+\mc H\right) (\mc Y_1^b)^{-1}.
\end{equation}
Here, $\mc H$ is contained in the set
\begin{equation}\label{eq:SH}
	S_{\mc H}=\{\mc H\in \mathbb R^{p_2\times m_b}:\mc H^T\mc H \leq Q_b\}.
\end{equation}
If $(A_{00},A_{00}A_{03}+A_{02})$ is controllable, we can always find $\mc H$ such that $(A_{00},K_{20})$ is observable. This is proved in the following. According to Lemma \ref{thm:eigenvector test for ctrl}, the controllability of $(A_{00},A_{00}A_{03}+A_{02})$ implies that no eigenvector of $diag\left(( A_{00}^a)^T, (A_{00}^b)^T\right)$ is in the kernel of  $\left[\begin{matrix}( A_{03}^a)^T( A_{00}^a)^T+( A_{02}^a)^T & ( A_{03}^b)^T( A_{00}^b)^T+( A_{02}^b)^T\end{matrix}\right]$. This implies that both $( A_{00}^a,  A_{00}^a A_{03}^a+ A_{02}^a)$ and $( A_{00}^b,  A_{00}^b A_{03}^b+ A_{02}^b)$ are controllable, which can be proved by applying the eigenvector tests in Lemma \ref{thm:eigenvector test for ctrl} to the vectors $\small\left[\begin{matrix}\eta_a \\ 0\end{matrix}\right]$ and $\small\left[\begin{matrix}0 \\ \eta_b\end{matrix}\right]$, where $\eta_a$ and $\eta_b$ are eigenvectors of $( A_{00}^a)^T$ and $( A_{00}^b)^T$, respectively. According to Lemma \ref{thm:eigenvector test for obsv}, $(A_{00}, K_{20})$ is observable if and only if for any non-zero vector $\small\delta_K=\left[\begin{matrix}\delta_a \\ \delta_b\end{matrix}\right]$, which is an eigenvector of $ A_{00}$, we have $ K_{20} \delta_K\neq 0$. Since $ A_{00}^a$ and $ A_{00}^b$ have no common eigenvalues, then $\delta_K$ is an eigenvector of $ A_{00}$ only if $\delta_a=0$ or $\delta_b=0$. We consider two cases:

\textbf{Case 1}. $\delta_a\neq 0$ and $\delta_b=0$. In this case, $\delta_a$ is an eigenvector of $ A_{00}^a$; i.e., $ A_{00}^a\delta_a=\lambda_a \delta_a$ for some scalar $\lambda_a$. Since $ A_{00}^a+( A_{00}^a)^T=0$, we have $( A_{00}^a)^T \delta_a=-\lambda_a \delta_a$. Hence, $(A_{00}^a)^{-T}\delta_a=-\frac{1}{ \lambda_a}\delta_a$. Also, because $(A_{00}^a,  A_{00}^a A_{03}^a+ A_{02}^a)$ is controllable, $\left(( A_{03}^a)^T( A_{00}^a)^T+( A_{02}^a)^T\right)\delta_a\neq 0$. Therefore,
\begin{align*}
	 K_{20}\delta_K= K_{20}^a\delta_a=& \left(-( A_{02}^a)^T( A_{00}^a)^{-T}-( A_{03}^a)^T\right)\delta_a\notag\\
	=&-\left(( A_{03}^a)^T( A_{00}^a)^T+( A_{02}^a)^T\right)(A_{00}^a)^{-T}\delta_a\notag\\
	=&\ \frac{1}{ \lambda_a}\left(( A_{03}^a)^T( A_{00}^a)^T+( A_{02}^a)^T\right)\delta_a\neq 0.
\end{align*}

\textbf{Case 2}. $\delta_a=0$ and $\delta_b \neq 0$. In this case, $\delta_b$ is an eigenvector of $A_{00}^b$.
Because $-( A_{02}^b)^T( A_{00}^b)^{-T}-( A_{03}^b)^T$ in (\ref{eq:K20b}) is fixed, and $S_{\mc H}$ has nonempty interior due to the positive definiteness of $Q_b$, then we can always find $\mc H$ such that
\begin{align*}
 K_{20}\delta_K =&\  K_{20}^b\delta_b \notag\\
 =& \left(-( A_{02}^b)^T( A_{00}^b)^{-T}-( A_{03}^b)^T+ \mc H\right)({\mc Y}_1^b)^{-1}\delta_b\notag\\
\neq &\ 0,
\end{align*}
for all $\delta_b$ that are eigenvalues of $A_{00}^b$. We conclude that there exists $\mc H$ such that $(A_{00}, K_{20})$ is observable. We will choose such a matrix $ \mc H$ in the following proof.

Let $K_{10}$ be defined as
\begin{equation}\label{eq:rd1 K_10}
	K_{10}= \left[\begin{matrix}K_{10}^a & K_{10}^b\end{matrix}\right],
\end{equation}
where
\begin{equation}\label{eq:K10a}
	K_{10}^a=-{A_{01}^a}^T(A_{00}^a)^{-T},
\end{equation}
and
\begin{equation}\label{eq:K10b}
	K_{10}^b=\left(-{A_{01}^b}^T(A_{00}^b)^{-T}-K_{13}\mc H\right) (\mc Y_1^b)^{-1}.
\end{equation}
Here $K_{13}$ is contained in the set
\begin{equation}\label{eq:K13}
	S_K=\{K_{13}\in \mathbb R^{p_1\times p_2}:K_{13}K_{13}^T \leq 2I\}.
\end{equation}
If $(A_{00},A_{01})$ is controllable, we can always find $K_{13}$ such that $(A_{00},K_{10})$ is observable. This is proved in the following. According to Lemma \ref{thm:eigenvector test for ctrl}, the controllability of $(A_{00}, A_{01})$ implies that no eigenvector of $diag\left((A_{00}^a)^T,(A_{00}^b)^T\right)$ is in $ker\left(\left[\begin{matrix}(A_{01}^a)^T & (A_{01}^b)^T\end{matrix}\right]\right)$. This implies that both $(A_{00}^a,A_{01}^a)$ and $(A_{00}^b,A_{01}^b)$ are controllable, which can be proved by applying the eigenvector tests in Lemma \ref{thm:eigenvector test for ctrl} to the vectors $\small\left[\begin{matrix}\eta_a\\ 0\end{matrix}\right]$ and $\small\left[\begin{matrix}0\\ \eta_b\end{matrix}\right]$, where $\eta_a$ and $\eta_b$ are eigenvectors of $(A_{00}^a)^T$ and $(A_{00}^b)^T$, respectively. According to Lemma \ref{thm:eigenvector test for obsv}, $(A_{00},K_{10})$ is observable if and only if for any non-zero vector $\small\phi_K = \left[\begin{matrix}\phi_a \\ \phi_b\end{matrix}\right]$, which is an eigenvector of $A_{00}$, we have that $K_{10}\phi_K\neq 0$. Since $A_{00}^a$ and $A_{00}^b$ have no common eigenvalues, then $\phi_K$ is an eigenvector of $A_{00}$ only if $\phi_a = 0$ or $\phi_b=0$. We consider two cases:

\textbf{Case 1}. $\phi_a \neq 0$ and $\phi_b = 0$. In this case, $\phi_a$ is an eigenvector of $A_{00}^a$; i.e., $A_{00}^a \phi_a = \mu_a \phi_a$ for some scalar $\mu_a$. Since $A_{00}^a+(A_{00}^a)^T=0$, we have $(A_{00}^a)^T\phi_a = -\mu_a \phi_a$. Hence, $(A_{00}^a)^{-T}\phi_a = -\frac{1}{\mu_a} \phi_a$. Also, because $(A_{00}^a,A_{01}^a)$ is controllable, $(A_{01}^a)^T\phi_a\neq 0$. Therefore,
\begin{equation*}
K_{10}\phi_K=K_{10}^a\phi_a=-(A_{01}^a)^T(A_{00}^a)^{-T} \phi_a = \frac{1}{\mu_a}(A_{01}^a)^T	\phi_a \neq 0.
\end{equation*}

\textbf{Case 2}. $\phi_a = 0$ and $\phi_b \neq 0$. In this case, $\phi_b$ is an eigenvector of $A_{00}^b$. Because $-(A_{01}^b)^T(A_{00}^b)^{-T}$ is fixed and the set $S_K$ has a nonempty interior, then we can always find $K_{13}$, together with an $\mc H$ that makes $(A_{00}, K_{20})$ observable, such that
\begin{align*}
K_{13}\phi_K=&K_{13}^b\phi_b \\
=&\left(-{A_{01}^b}^T\left(A_{00}^b\right)^{-T} - K_{13}\mc H\right) (\mc Y_1^b)^{-1}\phi_b \neq 0,
\end{align*}
for all $\phi_b$ that are eigenvectors of $A_{00}^b$. Therefore, with this particular choice of $K_{13}$, we have that $(A_{00},K_{10})$ is observable.

Now recall that $(A,B)$ is controllable if and only if $(A_{00},A_{01})$ or $(A_{00},A_{00}A_{03}+A_{02})$ is controllable. Also, $(A,C)$ is observable if and only if $(A_{00},K_{10})$ or $(A_{00},K_{20})$ is observable. Since the controllability of $(A_{00},A_{01})$ implies the observability of $(A_{00},K_{10})$ and the controllability of $(A_{00},A_{00}A_{03}+A_{02})$ implies the observability of $(A_{00},K_{20})$, then the controllability of $(A,B)$ implies the observability of $(A,C)$, when suitable $\mc H$ and $K_{13}$ are chosen.

Therefore, with those choices of $\mc H$ and $K_{13}$, the controllability of the system (\ref{eq:original normal}) implies that the realisation $(A,B,C)$ in (\ref{eq:rd1 A}), (\ref{eq:rd1 B}) and (\ref{eq:rd1 C}) is minimal.

Choose the other state feedback matrices as follows:
\begin{align}
K_{11}=& K_{10}A_{00}^{-1}A_{01}-\mc Y_2^{-1},\label{eq:K_11}\\
K_{12}=& K_{10}A_{00}^{-1}A_{02},\label{eq:K_12}\\
K_{21} =& K_{20}A_{00}^{-1}A_{01},\label{eq:K_21}\\
K_{22} =& K_{20}A_{00}^{-1}A_{02}-\mc Y_3^{-1},\label{eq:K_22}\\
K_{23} =& -\frac{1}{2} I,\label{eq:K_23}
\end{align}
where $\mc Y_2 \in \mathbb R^{p_1\times p_1}$ and $\mc Y_3 \in \mathbb R^{p_2\times p_2}$ can be any symmetric positive definite matrices; i.e., $\mc Y_2=\mc Y_2^T >0$ and $\mc Y_3 = \mc Y_3^T>0$. We will apply Lemma \ref{lemma:NI} in the following in order to prove that the system (\ref{eq:linearised2}) is an NI system. We construct the matrix $Y$ as follows:
\begin{equation}\label{eq:Y}
Y=\left[\begin{matrix}Y_{11} & -A_{00}^{-1}A_{01}\mc Y_2 & -A_{00}^{-1}A_{02}\mc Y_3 & 0\\-\mc Y_2 A_{01}^TA_{00}^{-T} & \mc Y_2 & 0 & 0 \\ -\mc Y_3 A_{02}^TA_{00}^{-T} & 0 & \mc Y_3 & 0 \\ 0 & 0 & 0 & I\end{matrix}\right],
\end{equation}
where $Y_{11}=\mc Y_1 + A_{00}^{-1}A_{01}\mc Y_2 A_{01}^TA_{00}^{-T}+A_{00}^{-1}A_{02}\mc Y_3 A_{02}^TA_{00}^{-T}$. Here, $\mc Y_1=diag(y_1^a I,\mc Y_1^b)$ with $y_1^a>0$ being a scalar. It can be verified that $Y>0$ using the Schur complement theorem.

In order to verify Condition 1 in Lemma \ref{lemma:NI}, we note that for the determinant of the matrix $A$ in (\ref{eq:rd1 A}) we have
\begin{align*}
	&(-1)^{p_2}\det{A}=\det\left[\begin{matrix}A_{00}&A_{01}&A_{02}&A_{03}\\K_{10}&K_{11} & K_{12} & K_{13}\\ K_{20} & K_{21} & K_{22} & K_{23} \\ 0&0&0&I \end{matrix}\right]\\
	=& \det\left[\begin{matrix}A_{00}&A_{01}&A_{02}\\K_{10}&K_{11} & K_{12}\\ K_{20} & K_{21} & K_{22}\end{matrix}\right]\\
	=&\det{A_{00}}\det{\left(\left[\begin{matrix}K_{11} & K_{12}\\ K_{21} & K_{22}\end{matrix}\right]-\left[\begin{matrix}K_{10}\\ K_{20}\end{matrix}\right]A_{00}^{-1}\left[\begin{matrix}A_{01} & A_{02}\end{matrix}\right]\right)}\\
	=&\det{A_{00}}\det{\left[\begin{matrix}-\mc Y_2^{-1} & 0 \\ 0 & -\mc Y_3^{-1}\end{matrix}\right]}\\
	=&\det A_{00}\det(-\mc Y_2^{-1}) \det(-\mc Y_3^{-1})\\
	\neq & \ 0,
\end{align*}
where the equalities also use (\ref{eq:K_11})-(\ref{eq:K_22}). Also, the input feedthrough matrix in the system (\ref{eq:linearised2}) is zero, and hence symmetric. Hence, Condition 1 in Lemma \ref{lemma:NI} is satisfied. For Condition 2 in Lemma \ref{lemma:NI}, with $Y$ defined in (\ref{eq:Y}), we have
\begin{equation}\label{eq:AY}
AY=	\left[\begin{matrix}y_1^a A_{00}^a & 0 & 0 & 0 & A_{03}^a\\ 0 & A_{00}^b\mc Y_1^b & 0 & 0 & A_{03}^b \\0 & -K_{13}\mc H & -I & 0 & K_{13} \\ 0 & 0&0&0& I \\ -(A_{03}^a)^T & \mc H-(A_{03}^b)^T & 0 & -I & K_{23}\end{matrix}\right].
\end{equation}
Therefore, we have that
\begin{equation}\label{eq:AYCT=-B}
	AYC^T=-B,
\end{equation}
and
\begin{equation*}
AY+YA^T =\left[\begin{matrix}0 & 0 & 0 & 0 & 0\\ 0 & -\mc Q_b & -\mc H^TK_{13}^T & 0  & \mc H^T  \\0 & -K_{13}\mc H & -2I & 0 & K_{13} \\ 0 & 0&0&0& 0 \\ 0 & \mc H & K_{13}^T & 0 & K_{23}+K_{23}^T\end{matrix}\right].
\end{equation*}
Let
\begin{align*}
M =& \left[\begin{matrix} -\mc Q_b & -\mc H^TK_{13}^T & \mc H^T  \\ -K_{13}\mc H & -2I & K_{13} \\ \mc H & K_{13}^T & K_{23}+K_{23}^T\end{matrix}\right]\notag\\
=&\left[\begin{matrix} -\mc Q_b & -\mc H^TK_{13}^T & \mc H^T  \\ -K_{13}\mc H & -2I & K_{13} \\ \mc H & K_{13}^T & -I\end{matrix}\right],
\end{align*}
where (\ref{eq:K_23}) is used. For the matrix $-M$, we have that $I>0$ and the Schur complement of the block $I$ is
\begin{align*}
(-M)/I =& \left[\begin{matrix} \mc Q_b & \mc H^TK_{13}^T\\ K_{13}\mc H & 2I \end{matrix}\right]-\left[\begin{matrix} \mc H^T \\ K_{13}\end{matrix}\right]\left[\begin{matrix} \mc H & K_{13}^T\end{matrix}\right]\notag\\
=&\left[\begin{matrix} \mc Q_b - \mc H^T\mc H & 0 \\ 0 & 2I-K_{13}K_{13}^T \end{matrix}\right]\geq 0,
\end{align*}
where (\ref{eq:SH}) and (\ref{eq:K13}) are also used. Therefore, we have that $M \leq 0$. Hence $AY+YA^T \leq 0$, and Condition 2 in Lemma \ref{lemma:NI} is satisfied. Hence, the system (\ref{eq:linearised2}) is an NI system with a minimal realisation.

\textbf{Necessity}. If the realisation $(A,B,C)$ in (\ref{eq:rd1 A}), (\ref{eq:rd1 B}) and (\ref{eq:rd1 C}) is minimal and NI, then according to the proof of Lemma \ref{lemma:NI} (see Lemma 7 in \cite{xiong2010negative}), there exists an $X=X^T > 0$ such that
\begin{equation*}
	\left[\begin{matrix} XA+A^TX & XB-A^TC^T \\ B^TX-CA & -(CB+B^TC^T)\end{matrix}\right] \leq 0.
\end{equation*}
Therefore, for any $z$, $x_1$, $x_2$, $x_3$ and $v$, we have
\begin{equation}\label{eq:rd1 LMI z y v}
	\left[\begin{matrix} z \\ x_1 \\ x_2 \\ x_3 \\ v\end{matrix}\right]^T\left[\begin{matrix} XA+A^TX & XB-A^TC^T \\ B^TX-CA & -(CB+B^TC^T)\end{matrix}\right]\left[\begin{matrix} z \\ x_1 \\ x_2 \\ x_3 \\ v\end{matrix}\right] \leq 0.
\end{equation}
Let $\small X=\left[\begin{matrix} X_{11} & X_{12} & X_{13} & X_{14}\\ X_{12}^T & X_{22} & X_{23} & X_{24} \\  X_{13}^T & X_{23}^T & X_{33} & X_{34} \\ X_{14}^T & X_{24}^T & X_{34}^T & X_{44}\end{matrix}\right]$ and substitute (\ref{eq:rd1 A}), (\ref{eq:rd1 B}) and (\ref{eq:rd1 C}) into (\ref{eq:rd1 LMI z y v}). Also, take $x_1=0$, $x_2 = x_3 = 0$ and $\tilde v=\left[\begin{matrix} -K_{10} \\ -K_{20} \end{matrix}\right]z$. We get
\begin{equation*}
	z^T(X_{11}A_{00}+A_{00}^TX_{11})z \leq 0
\end{equation*}
for any $z$, which implies that $X_{11}A_{00}+A_{00}^TX_{11}\leq 0$. Considering $X=X^T>0$, we have $X_{11}>0$. Also, since $\det A_{00}\neq 0$, then according to Lemma \ref{thm:marginally stable}, $A_{00}$ is Lyapunov stable. This completes the proof.
\end{IEEEproof}

To facilitate the description of the necessary and sufficient conditions for state feedback equivalence to a system in the general form (\ref{eq:original}), we recall the following terminology (see \cite{isidori2013nonlinear,khalil2002nonlinear}). In the case when the system (\ref{eq:output transformed system}) has relative degree less than or equal to two, the system (\ref{eq:original normal}) is said to be the \emph{normal form} of (\ref{eq:output transformed system}). The dynamics (\ref{eq:internal}), which are not controlled by the input $u$ directly or through chains of integrators, are called the \emph{internal dynamics}. The other part of the state, described by (\ref{eq:rd1})-(\ref{eq:rd2d}), are called the \emph{external dynamics}. Setting the states described by the external dynamics to be zero in the internal dynamics, we obtain the \emph{zero dynamics}:
\begin{equation}\label{eq:zero dynamics}
	\dot z = A_{00}z.
\end{equation}
We now provide the definition of the weakly minimum phase property.

\begin{definition}(Weakly Minimum Phase)\cite{saberi1990global,byrnes1991passivity}
The system (\ref{eq:output transformed system}) of relative degree less than or equal to two is said to be weakly minimum phase if its zero dynamics (\ref{eq:zero dynamics}) are Lyapunov stable.
\end{definition}

\begin{theorem}\label{thm:feedback NI}
Suppose the system (\ref{eq:original}) satisfying $rank(\mc B)=rank(\mc C)=p$ is minimal with no zero at the origin. Then it is state feedback equivalent to an NI system if and only if there exists an output transformation $\tilde y = T_y  y$, where $T_y \in \mathbb R^{p\times p}$ and $\det T_y \neq 0$, such that the transformed system has relative degree less than or equal to two, and the transformed system is weakly minimum phase.
\end{theorem}
\begin{IEEEproof}
	\textbf{Sufficiency.} The sufficiency part of the proof directly follows from Lemmas \ref{lem:feedback NI irrelevance}, \ref{lemma:normal form} and \ref{lem:feedback NI}. According to Lemma \ref{lemma:normal form}, the system (\ref{eq:original}) can always be transformed into the form (\ref{eq:original normal}) using nonsingular input, output and state transformations. Since the system (\ref{eq:original}) has no zero at the origin, then $\det A_{00} \neq 0$ because nonsingular input, output and state transformations do not change the zeros of a system. Also, since the transformed system is weakly minimum phase, then $A_{00}$ is Lyapunov stable. Since the input, output and state transformations are all nonsingular, the minimality of the system (\ref{eq:original}) is preserved in (\ref{eq:original normal}). According to Lemma \ref{lem:feedback NI}, the output transformed system (\ref{eq:original normal}) is state feedback equivalent to an NI system. According to Lemma \ref{lem:feedback NI irrelevance}, the original system (\ref{eq:original}) is also state feedback equivalent to an NI system. This completes the sufficiency part of the proof.

\textbf{Necessity.} We first prove that if the system (\ref{eq:original}) is state feedback equivalent to an NI system, then there exists an output transformation $\tilde y = T_y y$ that transforms the system (\ref{eq:original}) into a system with relative degree less than or equal to two. 

If the system (\ref{eq:original}) is state feedback equivalent to an NI system, then according to Lemma \ref{lemma:NI property preservation}, it is still feedback equivalent NI after a nonsingular output transformation. We apply an output transformation to the system $(\mc A, \mc B,\mc C)$ in order that the transformed system has a leading incomplete relative degree vector. Since the output transformed system is feedback equivalent to an NI system, then under a state feedback control law, we can make it NI with a minimal realization $(\hat A,\hat B,\hat C)$.

Since the system with realization $(\hat A, \hat B, \hat C)$ has a leading incomplete relative degree vector $r$, we denote by $p_1 \geq 0$ the number of components in $r$ that equal to one; i.e., $r_1,\cdots,r_{p_1} = 1$, and $r_{p_1+1},\cdots,r_{p}\geq 2$. We decompose the matrix $\hat C$ as
\begin{equation*}
\hat C = \left[\begin{matrix} \hat C_O \\ \hat C_G\end{matrix}\right],
\end{equation*}
where $\hat C_O \in \mathbb R^{p_1 \times n}$ and $\hat C_G \in \mathbb R^{(p-p_1) \times n}$. Here, $\hat C_O$ determines the output entries corresponding to the ones in $r$, and $\hat C_G$ determines the output entries corresponding to the components greater than one in $r$. According to Definition \ref{def:LIRD}, $rank(\hat C_O \hat B) =p_1$ and $\hat C_G \hat B = 0 $.

According to the proof of Lemma \ref{lemma:NI} (see \cite{xiong2010negative}), the fact that $(\hat A,\hat B,\hat C)$ is NI implies that there exists $X=X^T>0$ such that
\begin{equation}\label{eq:NI ineq hat}
\left[\begin{matrix} X\hat A+\hat A^TX & X\hat B-\hat A^T\hat C^T \\ \hat B^TX-\hat C\hat A & -(\hat C\hat B+\hat B^T\hat C^T)\end{matrix}\right] \leq 0.	
\end{equation}
Decomposing $\hat B$ accordingly as $\hat B = \left[\begin{matrix} \hat B_O & \hat B_G\end{matrix}\right]$ where $\hat B_O \in \mathbb R^{n\times p_1}$ and $\hat B_G \in \mathbb R^{n\times (p-p_1)}$, the inequality (\ref{eq:NI ineq hat}) can be expanded to be
\begin{equation}\label{eq:NI ineq hat expanded}
\small\left[\begin{matrix} X\hat A+\hat A^TX & X\hat B_O-\hat A^T\hat C_O^T & X\hat B_G-\hat A^T\hat C_G^T \\ \hat B_O^TX-\hat C_O\hat A & -(\hat C_O\hat B_O+\hat B_O^T\hat C_O^T) & -\hat C_O \hat B_G \\ \hat B_G^TX-\hat C_G\hat A & -\hat B_G^T\hat C_O^T & 0\end{matrix}\right] \leq 0,	
\end{equation}
where the condition $\hat C_G \hat B = 0$ is also used. The condition (\ref{eq:NI ineq hat expanded}) implies that $\hat C_O\hat B_G = 0$ and $\hat B_G^TX-\hat C_G\hat A = 0$. We have that $rank(\hat B_G)=p-p_1$ because $rank(\hat B)=rank(\mc B)=p$. Then, $\hat B_G^TX-\hat C_G\hat A = 0$ implies that $\hat C_G\hat A \hat B_G = \hat B_G^TX \hat B_G > 0$. The positive definiteness of $\hat C_G \hat A \hat B_G$ implies that the largest component in the leading incomplete relative degree vector $r$ of the system is two. Moreover, we have that
\begin{equation}\label{eq:H(r) in thm 1}
	\left[\begin{matrix} \hat C_O\hat B \\ \hat C_G \hat A \hat B \end{matrix}\right] = \left[\begin{matrix} \hat C_O\hat B_O & 0 \\ \hat C_G \hat A \hat B_O & \hat C_G \hat A \hat B_G \end{matrix}\right].
\end{equation}
Since $\hat C_O \hat B_G =0$ and $rank(\hat C_O \hat B)=p_1$, we have that $\det(\hat C_O \hat B_O) \neq 0$. Considering that $\hat C_O \hat B_O$ and $\hat C_G \hat A \hat B_G$ in (\ref{eq:H(r) in thm 1}) are both nonsingular, we have that $\det\left[\begin{matrix} \hat C_O\hat B \\ \hat C_G \hat A \hat B \end{matrix}\right] \neq 0$. This implies that the leading incomplete relative degree vector $r$ of the realization $(\hat A ,\hat B,\hat C)$ is indeed a relative degree vector, whose components are either one or two. Therefore, we conclude that the system (\ref{eq:original}) can be output transformed into a system with a relative degree vector $r=\{r_1,\cdots,r_p\}$ with $1 \leq r_i\leq 2$ for all $i=1,\cdots,p$. Therefore, the system (\ref{eq:original}) can be transformed into the form (\ref{eq:original normal}) using input, output and state transformations. The necessity part of Lemma \ref{lem:feedback NI} implies that the weakly minimum phase property of the output transformed system is another necessary condition. This completes the necessity part of the proof.
\end{IEEEproof}

We also derive necessary and sufficient conditions under which the system (\ref{eq:original}) can be rendered OSNI.

\begin{definition}\label{def:state feedback equivalent OSNI}
A system in the form of (\ref{eq:original}) is said to be state feedback equivalent to an OSNI system if there exists a state feedback control law
\begin{equation*}
u=K_x x+ K_v v,	
\end{equation*}
where $K_x\in \mathbb R^{p\times n}$ and $K_v \in \mathbb R^{p\times p}$, such that the closed-loop system with the new input $v\in \mathbb R^p$ is minimal and OSNI.
\end{definition}

\begin{lemma}\label{lemma:OSNI property preservation}
If the transfer matrix $R(s)$ is OSNI, then $T R(s) T^T$ is also OSNI, where $T\in \mathbb R^{p\times p}$ and $\det T \neq 0$.
\end{lemma}
\begin{IEEEproof}
The proof follows from Definition \ref{def:OSNI}. If $R(s)$ is OSNI, then we have that
\begin{equation*}
j\omega [R(j\omega)-R(j\omega)^*]-\epsilon \omega^2 \bar R(j\omega)^*\bar R(j\omega) \geq 0,
\end{equation*}
$\forall \omega \in \mathbb R \cup {\infty}$ where $\bar R(j\omega) = R(j\omega)-R(\infty)$. We have that $T^TT \leq \lambda_{max}(T^TT)I$. Therefore,
\begin{align}
	& j\omega [R(j\omega)-(R(j\omega))^*]-\frac{\epsilon}{\lambda_{max}(T^TT)} \omega^2 \bar R(j\omega)^*T^TT\bar R(j\omega)\notag\\
	& \geq j\omega [R(j\omega)-(R(j\omega))^*]-\frac{\epsilon}{\lambda_{max}(T^TT)} \omega^2 \bar R(j\omega)^*T^TT\bar R(j\omega)\notag\\
	&+\frac{\epsilon}{\lambda_{max}(T^TT)} \omega^2 \bar R(j\omega)^*(T^TT-\lambda_{max}(T^TT)I)\bar R(j\omega)\notag\\
	&=j\omega [R(j\omega)-R(j\omega)^*]-\epsilon \omega^2 \bar R(j\omega)^*\bar R(j\omega) \geq 0\label{eq:OSNI derivation}.
\end{align}
The transfer matrix $T R(s) T^T$ satisfies Definition \ref{def:OSNI} via (\ref{eq:OSNI derivation}). Therefore, the transformed system $TR(s)T^T$ is OSNI with the output strictness $\frac{\epsilon}{\lambda_{max}(T^TT)} $.
\end{IEEEproof}

We show in the following that the same conditions in Theorem \ref{thm:feedback NI} also lead to state feedback equivalence to an OSNI system.
\begin{lemma}\label{lem:feedback OSNI}
Suppose the system (\ref{eq:original normal}) has $\det A_{00} \neq 0$. Then it is state feedback equivalent to an OSNI system if and only if it is controllable and $A_{00}$ is Lyapunov stable.
\end{lemma}
\begin{IEEEproof}
The necessity part this lemma follows from the necessity part of Lemma \ref{lem:feedback NI} because OSNI systems belong to the class of NI systems.

For the sufficiency part, we need to show that the condition
\begin{equation*}
AY+YA^T+\epsilon(CAY)^TCAY \leq 0	
\end{equation*}
in Lemma \ref{lem:OSNI} is satisfied for some scalar $\epsilon>0$ in addition to what is shown in the sufficiency proof of Lemma \ref{lem:feedback NI}. Following from the sufficiency proof of Lemma \ref{lem:feedback NI}, we add a restriction on the choice of $K_{13}$ such that $K_{13}^TK_{13}=I$. Note that this additional restriction does not change the results in Lemma \ref{lem:feedback NI}. Using $C$ and $AY$ in (\ref{eq:new C}) and (\ref{eq:AY}), we have that
\begin{equation*}
CAY = \left[\begin{matrix}0& -K_{13}\mc H&-I&0&K_{13}\\0&0&0&0&I	
\end{matrix}
\right].
\end{equation*}
Therefore,
\begin{equation*}
(CAY)^TCAY = \left[\begin{matrix}0&0&0&0&0\\0& \mc H^T \mc H & \mc H^TK_{13}^T & 0 & -\mc H^T \\ 0 & K_{13}\mc H & I & 0 & -K_{13} \\ 0 & 0&0&0&0 \\ 0& -\mc H & -K_{13}^T & 0& 2I	
\end{matrix}
\right].
\end{equation*}
Hence,
\begin{align*}
&AY+YA^T+ \epsilon (CAY)^TCAY \notag\\
=&\left[\begin{matrix}0 & 0 & 0 & 0 & 0\\ 0 & -\mc Q_b+\epsilon \mc H^T \mc H & -(1-\epsilon)\mc H^TK_{13}^T & 0  & (1-\epsilon)\mc H^T  \\0 & -(1-\epsilon) K_{13}\mc H & -(2-\epsilon)I & 0 & (1-\epsilon)K_{13} \\ 0 & 0&0&0& 0 \\ 0 & (1-\epsilon)\mc H & -(1-\epsilon) K_{13}^T & 0 & (1-2\epsilon)I\end{matrix}\right].
\end{align*}
Let
\begin{equation*}
\tilde M =\left[\begin{matrix} -\mc Q_b+\epsilon \mc H^T \mc H & -(1-\epsilon)\mc H^TK_{13}^T  & (1-\epsilon)\mc H^T  \\ -(1-\epsilon) K_{13}\mc H & -(2-\epsilon)I  & (1-\epsilon)K_{13} \\ (1-\epsilon)\mc H & (1-\epsilon) K_{13}^T & -(1-2\epsilon)I\end{matrix}\right].
\end{equation*}
We apply the Schur complement theorem in the following to find the range of $\epsilon$.
We choose $\epsilon \in (0,\frac{1}{2})$ and therefore $-(1-2\epsilon)I<0$. The Schur complement of the block $(1-2\epsilon)$ of $-\tilde M$ is
\begin{align*}
&(-\tilde M)/[(1-2\epsilon)I]	\notag\\
=&\left[\begin{matrix}
\mc Q_b-\epsilon \mc H^T \mc H & (1-\epsilon)\mc H^T K_{13}^T \\ (1-\epsilon) K_{13} \mc H & (2-\epsilon)I
\end{matrix}\right]\notag\\
&-\frac{(1-\epsilon)^4}{1-2\epsilon}\left[\begin{matrix}
\mc H^T \\ K_{13}
\end{matrix}\right]\left[\begin{matrix}
\mc H & K_{13}^T
\end{matrix}\right]\notag\\
=&\left[\begin{matrix}
\mc Q_b-\left(\epsilon+\frac{(1-\epsilon)^4}{1-2\epsilon}\right) \mc H^T \mc H & \left(1-\epsilon-\frac{(1-\epsilon)^4}{1-2\epsilon}\right)\mc H^T K_{13}^T \\ \left(1-\epsilon-\frac{(1-\epsilon)^4}{1-2\epsilon}\right) K_{13} \mc H & \left(2-\epsilon-\frac{(1-\epsilon)^4}{1-2\epsilon}\right)I
\end{matrix}\right],
\end{align*}
which is positive semi-definite when $\epsilon \in \left(0,\frac{1}{2}\left(3-\sqrt{5}\right)\right]$. In this case, $AY+YA^T+ \epsilon (CAY)^TCAY \leq 0$. Therefore, the system with the realization $(A,B,C)$ in (\ref{eq:rd1 A}), (\ref{eq:rd1 B}) and (\ref{eq:rd1 C}) is OSNI.
\end{IEEEproof}

\begin{theorem}\label{thm:feedback OSNI}
Suppose the system (\ref{eq:original}) is minimal with no zero at the origin. Then it is state feedback equivalent to an OSNI system if and only if there exists an output transformation $\tilde y = T_y y$, where $T_y \in \mathbb R^{p\times p}$ and $\det T_y \neq 0$, such that the transformed system has relative degree less than or equal to two, and the transformed system is weakly minimum phase.
\end{theorem}
\begin{IEEEproof}
This proof is similar to the proof of Theorem \ref{thm:feedback NI} except that Lemmas \ref{lemma:OSNI property preservation} and \ref{lem:feedback OSNI} are used instead of Lemmas \ref{lemma:NI property preservation} and \ref{lem:feedback NI}.
\end{IEEEproof}

Considering the results in Theorem \ref{thm:feedback NI} and \ref{thm:feedback OSNI}, we have the following corollary.
\begin{corollary}\label{corollary:feedback NI and OSNI}
Suppose the system (\ref{eq:original}) is minimal with no zero at the origin. Then the following statements are equivalent:

1. The system (\ref{eq:original}) is state feedback equivalent to an NI system;

2. The system (\ref{eq:original}) is state feedback equivalent to an OSNI system;

3. There exists an output transformation $\tilde y = T_y  y$, where $T_y \in \mathbb R^{p\times p}$ and $\det T_y \neq 0$, such that the transformed system has relative degree less than or equal to two, and the transformed system is weakly minimum phase.
\end{corollary}

\section{STATE FEEDBACK EQUIVALENCE TO AN SSNI SYSTEM}\label{section:SSNI}

In this section, we derive necessary and sufficient conditions under which a system in the form of (\ref{eq:original}) is state feedback equivalent to an SSNI system. First, we define state feedback equivalence to an SSNI system as follows.

\begin{definition}
A system in the form of (\ref{eq:original}) is said to be state feedback equivalent to an SSNI system if there exists a state feedback control law
\begin{equation*}
u=K_x x+ K_v v,	
\end{equation*}
where $K_x\in \mathbb R^{p\times n}$ and $K_v \in \mathbb R^{p\times p}$, such that the closed-loop system with the new input $v\in \mathbb R^p$ is SSNI.
\end{definition}

It will be shown later in this section that having a relative degree vector $r=\{1,\cdots,1\}$ is one of the necessary conditions for this system to be state feedback equivalent to an SSNI system. Therefore, we start with the derivation of the normal form for the system (\ref{eq:original}) with a relative degree vector $r=\{1,\cdots,1\}$.

\begin{lemma}\label{lemma:rd1 state transformation}
	Suppose the system (\ref{eq:original}) satisfying $rank(\mc B)=rank(\mc C) =p$ has a relative degree vector $r=\{1,\cdots,1\}$. Then there exists input and state transformations such that the resulting transformed system is of the form
\begin{subequations}\label{eq:rd1 only normal}
\begin{align}
\dot z =&\ A_{00}z+A_{01}y,\label{eq:rd1 internal}\\
\dot x_1 =&\ A_{10}z+A_{11}x_1+ \tilde u,\label{eq:linearised1b}\\
y=& \left[\begin{matrix}0& {I}\end{matrix}\right]\left[\begin{matrix}z \\ x_1\end{matrix}\right].\label{eq:linearised1c}
\end{align}
\end{subequations}
\end{lemma}
\begin{IEEEproof}
If (\ref{eq:original}) has a relative degree vector $r=\{1,\cdots,1\}$, then $\det(\mc C \mc B) \neq 0$. The rest of the proof follows from Lemma \ref{lemma:normal form} with $p_1 = p$ and $p_2=0$.
\end{IEEEproof}

Choose the input $u$ to be
\begin{equation*}
\tilde u=\left(v+(K_1-A_{10})z+(K_2-A_{11})y\right),
\end{equation*}
and the system (\ref{eq:rd1 only normal}) takes the form
\begin{subequations}\label{eq:rd1 only K}
\begin{align}
\dot z =&\ A_{00}z+A_{01}y,\label{eq:linearised2a}\\
\dot y =&\ K_1z+K_2y+v,\label{eq:linearised2b}\\
y=&\ [\begin{matrix}0& I\end{matrix}]\left[\begin{matrix}z \\ y\end{matrix}\right].\label{eq:linearised2c}
\end{align}
\end{subequations}
We need to find the state feedback matrices $K_1\in \mathbb R^{p\times m}$ and $K_2 \in \mathbb R^{p\times p}$ such that the system (\ref{eq:rd1 only K}) is SSNI.

\begin{lemma}\label{lem:SSNI}\cite{shi2021negative}
Suppose the system (\ref{eq:rd1 only K}) has $(A_{00},A_{01})$ controllable. Then the following statements are equivalent:

1. $A_{00}$ is Hurwitz;

2. There exist $K_1$ and $K_2$ such that the system (\ref{eq:rd1 only K}) is an SSNI system with realisation $(A,B,C)$, where $ A$ is Hurwitz, and the transfer function $R(s):=C(sI-  A)^{-1}B$ is such that $R(s)+R(-s)^T$ has full normal rank.
\end{lemma}
\begin{IEEEproof}
Let us define the following:
\begin{align}
A =& \left[\begin{matrix}A_{00} & A_{01} \\ K_1 & K_2\end{matrix}\right],\label{eq:SSNI A}\\
	B =& \left[\begin{matrix}0 \\ I\end{matrix}\right],\label{eq:SSNI B}\\
	C =& \left[\begin{matrix}0 & I\end{matrix}\right]\label{eq:SSNI C}.
\end{align}
From the proof of Lemma \ref{lem:feedback NI}, $(A,B)$ is controllable if and only if $(A_{00},A_{01})$ is controllable. Therefore, there are no observable uncontrollable modes in this system.

\textbf{Sufficiency}. Let $A_{00}$ be Hurwitz. Then according to Lemma \ref{thm:Lyapunov}, we can always find a matrix ${\mc Y}_1>0$ such that
\begin{equation*}
A_{00}{\mc Y}_1 + {\mc Y}_1 A_{00}^T + \frac{1}{2}A_{00}^{-1}A_{01}A_{01}^TA_{00}^{-T}<0	
\end{equation*}
is satisfied. In the sequel, we will find a matrix $ K_1$ such that
\begin{equation}\label{eq:rd1 SSNI Lyapunov ineq}
	A_{00}{\mc Y}_1 + {\mc Y}_1 A_{00}^T + \frac{1}{2}({\mc Y}_1 K_1^T+A_{00}^{-1}A_{01})( K_1{\mc Y}_1+A_{01}^TA_{00}^{-T})<0
\end{equation}
is satisfied. One possible choice is $ K_1=-A_{01}^TA_{00}^{-T}{\mc Y}_1^{-1}$, which simplifies (\ref{eq:rd1 SSNI Lyapunov ineq}) to be $A_{00}{\mc Y}_1 + {\mc Y}_1 A_{00}^T<0$. Let $ K_2 =  K_1A_{00}^{-1}A_{01}-{\mc Y}_2^{-1}$, where ${\mc Y}_2 \in \mathbb R^{p\times p}$ can be any symmetric positive definite matrix; i.e., ${\mc Y}_2={\mc Y}_2^T >0$. We apply Lemma \ref{thm:SSNI} in the following to prove that the system (\ref{eq:rd1 only K}) is an SSNI system. We construct the matrix $ Y$ as follows:
\begin{equation*}
 Y=\left[\begin{matrix}{\mc Y}_1 + A_{00}^{-1}A_{01}{\mc Y}_2 A_{01}^TA_{00}^{-T} & -A_{00}^{-1}A_{01}{\mc Y}_2\\-{\mc Y}_2 A_{01}^TA_{00}^{-T} & {\mc Y}_2\end{matrix}\right].
\end{equation*}
We have $ Y>0$ because ${\mc Y}_2>0$ and the Schur complement of the block ${\mc Y}_2$ is $ {\mc Y}_1$, which is positive definite. Now, we have $B+A YC^T=0$ and
\begin{equation*}
 A Y+ Y A^T = \left[\begin{matrix}A_{00}{\mc Y}_1 + {\mc Y}_1 A_{00}^T & {\mc Y}_1  K_1^T+A_{00}^{-1}A_{01} \\  K_1{\mc Y}_1+A_{01}^TA_{00}^{-T} & -2I\end{matrix}\right].
\end{equation*}
We have $-2I<0$ and the Schur complement of the block $2I$ in the matrix $-( A Y+ Y A^T)$ is
\begin{align*}
	(-( A &  Y+ Y  A^T)) /(2I)\notag\\
	=& -A_{00}{\mc Y}_1 - {\mc Y}_1 A_{00}^T\notag\\
	 &- \frac{1}{2}({\mc Y}_1  K_1^T+A_{00}^{-1}A_{01})( K_1{\mc Y}_1+A_{01}^TA_{00}^{-T})>0,
\end{align*}
according to (\ref{eq:rd1 SSNI Lyapunov ineq}). Hence $ A Y+ Y A^T<0$. According to Lemma \ref{thm:Lyapunov}, $ A^T$ is Hurwitz. Therefore $ A$ is Hurwitz. Now we prove that $R(s)+R(-s)^T$ has full normal rank. For $ A$, $B$ and $C$ given by (\ref{eq:SSNI A})-(\ref{eq:SSNI C}), we have
\begin{align}
	R(s)=&\ C(sI- A)^{-1}B\notag\\
	=&\left[\begin{matrix} 0 & I\end{matrix}\right]\left[\begin{matrix} sI-A_{00} & -A_{01}\\ - K_1 & sI- K_2\end{matrix}\right]^{-1}\left[\begin{matrix} 0\\ I\end{matrix}\right]\notag\\
	=&\left(sI- K_1(sI-A_{00})^{-1}A_{01}- K_2\right)^{-1}.\label{eq:rank arbitrary s}
\end{align}
Substituting $s=0$ in (\ref{eq:rank arbitrary s}), we have
\begin{equation*}
R(0)=( K_1A_{00}^{-1}A_{01}- K_2)^{-1}={\mc Y}_2>0.
\end{equation*}
Hence $R(s)+R(-s)^T$ must have full normal rank. Therefore, according to Lemma \ref{thm:SSNI}, the system (\ref{eq:rd1 only K}) is SSNI.

\textbf{Necessity}. If $ A$ is Hurwitz, $R(s)+R(-s)^T$ has full normal rank and the system (\ref{eq:rd1 only K}) is SSNI, then according to Lemma \ref{thm:SSNI}, there exists a matrix $ Y= Y^T>0$ such that $B=- A YC^T$ and $ A Y+ Y A^T<0$.

Letting $ X= Y^{-1}$, then $ X= X^T>0$. Also letting $Q = -( A Y+ Y A^T)$, then we have $Q=Q^T>0$ and $ X A+ A^T X = - X Q  X <0$. Since $B=- A YC^T$, we have $CB+B^TC^T=-C A YC^T-C Y A^TC^T=C QC^T$. Also, $ XB- A^TC^T=- X A X^{-1}C^T- A^TC^T=-( X A+ A^T X) X^{-1}C^T= XQ X X^{-1}C^T= XQC^T$. Since $Q=Q^T>0$, let $H:=Q^{\frac{1}{2}}$. Hence $H=H^T>0$. We have
\begin{align}
	\left[\begin{matrix}  X A+ A^T X &  XB- A^TC^T \\ B^T X-C A & -(CB+B^TC^T)\end{matrix}\right]=& -\left[\begin{matrix} L^T \\ W^T\end{matrix}\right]\left[\begin{matrix} L & W\end{matrix}\right]\notag\\
	\leq &\ 0,\label{eq:rd1 LMI SSNI}
\end{align}
where $L=H  X$ and $W= -H C^T$. (\ref{eq:rd1 LMI SSNI}) implies that for any $z\in \mathbb R^{m}$, $y\in \mathbb R^{p}$ and $v\in \mathbb R^{p}$, we have
\begin{align}
	&\left[\begin{matrix} z^T & y^T & v^T\end{matrix}\right]\left[\begin{matrix}  X A+ A^T X &  XB- A^TC^T \\ B^T X-C A & -(CB+B^TC^T)\end{matrix}\right]\left[\begin{matrix} z \\ y \\ v\end{matrix}\right]\notag\\
	&=-\left[\begin{matrix} z^T & y^T & v^T\end{matrix}\right]\left[\begin{matrix} L^T \\ W^T\end{matrix}\right]\left[\begin{matrix} L & W\end{matrix}\right]\left[\begin{matrix} z \\ y \\ v\end{matrix}\right] \leq 0,\label{eq:rd1 LMI z y v SSNI}
\end{align}
where equality holds if and only if $\small\left[\begin{matrix} L & W\end{matrix}\right]\left[\begin{matrix} z \\ y \\ v\end{matrix}\right]=0$. That is $L\left[\begin{matrix} z \\ y\end{matrix}\right]+Wv=0$, which is equivalent to $\small H\left( X\left[\begin{matrix} z \\ y\end{matrix}\right]-C^Tv\right)=0$. Because $H>0$, this equation holds if and only if
\begin{equation}\label{eq:rd1 z y v eq 0}
	 X\left[\begin{matrix} z \\ y\end{matrix}\right]-C^Tv=0.
\end{equation}
Let $\small  X=\left[\begin{matrix}  X_{11} &  X_{12}\\  X_{12}^T &  X_{22}\end{matrix}\right]$ and choose $y=0$ and $v=- K_1z$. With $C$ given by (\ref{eq:SSNI C}), (\ref{eq:rd1 z y v eq 0}) becomes
\begin{equation*}
	\left[\begin{matrix}  X_{11} \\  X_{12}^T+ K_1\end{matrix}\right]z = 0,
\end{equation*}
which holds only if $ X_{11}z=0$. Since $ X= X^T>0$, $ X_{11}= X_{11}^T>0$. Hence $ X_{11}z=0\iff z=0$. This implies that with the choice $y=0$ and $v=- K_1z$, strict inequality holds in (\ref{eq:rd1 LMI z y v SSNI}) for all $z\neq 0$. Substituting (\ref{eq:SSNI A})-(\ref{eq:SSNI C}) together with $y=0$ and $v=- K_1z$ into (\ref{eq:rd1 LMI z y v SSNI}), we obtain
\begin{equation*}
z^T( X_{11}A_{00}+A_{00}^T X_{11})z<0	
\end{equation*}
for all $z\neq 0$. This implies that $ X_{11}A_{00}+A_{00}^T X_{11}<0$. Therefore, according to Lemma \ref{thm:Lyapunov}, $A_{00}$ is Hurwitz.
\end{IEEEproof}

\begin{definition}(Minimum Phase)\cite{byrnes1991passivity,khalil2002nonlinear}
A system (\ref{eq:original}) satisfying $rank(\mc B)=rank(\mc C)=p$ with relative degree vector $\{1,\cdots,1\}$ is said to be minimum phase if its zero dynamics $\dot z=A_{00}z$ are asymptotically stable.
\end{definition}

\begin{theorem}\label{thm:feedback SSNI}
Suppose the system (\ref{eq:original}) satisfying $rank(\mc B)=rank(\mc C)=p$ is minimal. Then the following statements are equivalent:

1. The system has a relative degree vector $r=\{1,\cdots,1\}$ and is minimum phase;

2. The system is state feedback equivalent to an SSNI system with realisation $(A,B,C)$, where $ A$ is Hurwitz, and the transfer function $R(s):=C(sI-  A)^{-1}B$ is such that $R(s)+R(-s)^T$ has full normal rank.
\end{theorem}
\begin{IEEEproof}
	The proof from Statement 1 to Statement 2 follows directly from Lemmas \ref{lemma:rd1 state transformation} and \ref{lem:SSNI}. Note that the minimum phase condition is equivalent to the condition that $A_{00}$ is Hurwitz in Lemma \ref{lem:SSNI}. Now we prove that Condition 2 implies that the system has a relative degree vector $r=\{1,\cdots,1\}$. SSNI systems form a subclass of all NI systems according to Definition \ref{def:SSNI}. The analysis in the necessity proof of Theorem \ref{thm:feedback NI} also holds for SSNI systems except that strict inequalities hold for both (\ref{eq:NI ineq hat}) and (\ref{eq:NI ineq hat expanded}), where this additional restriction comes from the strict inequality in Lemma \ref{thm:SSNI}. Strict inequality for (\ref{eq:NI ineq hat expanded}) holds only if the zero block matrix has zero dimension, which is true only if $p_2=0$. This implies that statement 2 is true only if the original system (\ref{eq:original}) with realization $(\mc A, \mc B ,\mc C)$ can be output transformed by a nonsingular matrix $T_y \in \mathbb R^{p\times p}$ into a system with a relative degree vector $r=\{1,\cdots,1\}$. According to Definition \ref{def:RD vector}, that is, the output transformed system satisfies $\det(\tilde {\mc C} \mc B) \neq 0 $, where $\tilde {\mc C} = T_y \mc C$. Since $\det T_y \neq 0$, we have that $\det(\mc C \mc B) \neq 0$. This means that the original system (\ref{eq:original}) itself is already in a form with a relative degree vector $r=\{1,\cdots,1\}$. Therefore, according to Lemma \ref{lemma:rd1 state transformation}, (\ref{eq:rd1 only normal}) is the normal form of the system (\ref{eq:original}). The rest of the proof follows directly from Lemma \ref{lem:SSNI}.
\end{IEEEproof}

\section{CONTROL OF SYSTEMS WITH SNI UNCERTAINTY}\label{section:synthesis}
\begin{figure}[h!]
\centering
\psfrag{delta}{$\Delta(s)$}
\psfrag{nominal}{\hspace{-0.05cm}\small Nominal}
\psfrag{plant}{\small Plant}
\psfrag{controller}{\small Controller}
\psfrag{closed-loop}{\small Closed-Loop}
\psfrag{w}{$w$}
\psfrag{x}{$x$}
\psfrag{y}{$y$}
\psfrag{u}{$u$}
\psfrag{R_s}{\small $R(s)$}
\psfrag{+}{\small$+$}
\includegraphics[width=8.5cm]{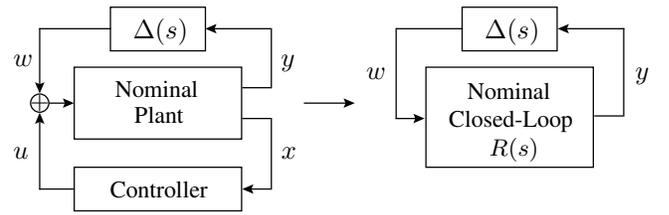}
\caption{A feedback control system. The plant uncertainty $\Delta(s)$ is SNI and satisfies $\lambda_{max}(\Delta(0))\leq \gamma$ and $\Delta(\infty)\geq 0$. Under some assumptions, we can find a controller such that the closed-loop transfer function $R(s)$ is NI with $R(\infty)=0$ and $\lambda_{max}(R(0))< 1/\gamma$. Then the closed-loop system is robustly stable.}
\label{fig:controller synthesis}
\end{figure}
One useful application of state feedback equivalence to NI systems is to robustly stabilize systems for a class of uncertainties. More precisely, for a system having SNI uncertainty, we can render the nominal closed-loop system NI with the DC gain condition satisfied when full state measurement is available. A similar controller synthesis problem is investigated in \cite{petersen2010feedback}, where the robust stabilzability depends on the solvability of a series of LMIs. However, in this paper, the LMI conditions in \cite{petersen2010feedback} are replaced by some simpler conditions with respect to the relative degree vector and the weakly minimum phase property.

Consider the uncertain feedback control system in Fig.~\ref{fig:controller synthesis} and suppose that full state feedback is available. Then Theorem \ref{thm:feedback NI} can be used in order to synthesize a state feedback controller such that the nominal closed-loop system is NI. Suppose the state-space model of the uncertain system in Fig.~\ref{fig:controller synthesis} is
\begin{subequations}\label{eq:original uncertain system}
\begin{align}
	\dot x =&\ \mc Ax+\mc B(u+w),\label{eq:uncertain A}\\
	y =&\ \mc C x,\label{eq:uncertain B}\\
	w =&\ \Delta y,\label{eq:uncertain SNI}
\end{align}	
\end{subequations}
where $x\in \mathbb R^n$, $u\in \mathbb R^{p}$ and $y\in \mathbb R^{p}$ are the state, input and output of the nominal plant. Here, (\ref{eq:uncertain SNI}) models the uncertainty, and the uncertainty transfer function $\Delta(s)$ is assumed to be SNI with $\Delta(\infty)\geq 0$ and $\lambda_{max}(\Delta(0))\leq \gamma$ for some constant $\gamma>0$.

The general idea used to stabilize the system (\ref{eq:original uncertain system}) is to choose a control law $u$ such that the system described by (\ref{eq:uncertain A}) and (\ref{eq:uncertain B}) is NI with input $w$ and output $y$. Therefore, since $\Delta(s)$ is SNI, the system (\ref{eq:original uncertain system}) forms a positive feedback interconnection of an NI system and an SNI system, whose equilibrium is asymptotically stable if 
the DC gain condition in Lemma \ref{lemma:dc gain theorem} is satisfied.

\begin{theorem}\label{thm:stabilization}
Consider the uncertain system (\ref{eq:original uncertain system}). Suppose the realization $(\mc A, \mc B,\mc C)$ is minimal with no zero at the origin. If there exists an output transformation $\tilde y = T_y y$, where $T_y \in \mathbb R^{p\times p}$ and $\det T_y \neq 0$, such that the realization $(\mc A,\mc B,T_y \mc C)$ has relative degree less than or equal to two and is weakly minimum phase, then there exist $K_x\in \mathbb R^{p\times n}$ and $K_w\in \mathbb R^{p\times p}$ such that the control law
\begin{equation*}
u=K_x x+ K_w w	
\end{equation*}
stabilizes the system (\ref{eq:original uncertain system}).
\end{theorem}
\begin{IEEEproof}
According to Theorem \ref{thm:feedback NI} and its proof, the conditions here imply that the nominal plant in (\ref{eq:original uncertain system}), described by
\begin{align*}
\dot x =&\ \mc Ax + \mc B u,\\
y =&\ \mc C x,
\end{align*}
is state feedback equivalent to an NI system. Suppose the corresponding state feedback control law is
\begin{equation*}
u = K_x x + K_v	v.
\end{equation*}
Therefore, the nominal plant, described by (\ref{eq:uncertain A}) and (\ref{eq:uncertain B}), is NI with input $w$ and output $y$ under the control law
\begin{equation*}
u = K_x x + K_w w,	
\end{equation*}
where $K_w = K_v - I$. Now the system (\ref{eq:original uncertain system}) is an interconnection of the nominal closed-loop NI system and the SNI uncertainty. To stabilize this interconnection, we investigate the DC gain conditions of Lemma \ref{lemma:dc gain theorem}. As is shown in the proof of Theorem \ref{thm:feedback NI}, the output transformed system $(\mc A,\mc B,T_y \mc C)$ is rendered NI with a transfer function $\hat R(s)$ where $\hat R(s)=C(sI-A)^{-1}B$ with $A$, $B$ and $C$ given by (\ref{eq:rd1 A}), (\ref{eq:rd1 B}) and (\ref{eq:rd1 C}). We have that $\hat R(\infty)=0$. With the state feedback matrices given in the proof of Theorem \ref{thm:feedback NI}, we also have that
\begin{equation*}
\hat R(0)=-CA^{-1}B=CA^{-1}AYC^T=CYC^T=\left[\begin{matrix}
\mc Y_2 & 0 \\ 0& \mc Y_3	
\end{matrix}
\right],	
\end{equation*}
where we also use (\ref{eq:Y}) and (\ref{eq:AYCT=-B}). Since the NI state feedback equivalence of the realization $(\mc A,\mc B,\mc C)$ follows from the NI state feedback equivalence of the output transformed system $(\mc A,\mc B,T_y \mc C)$, using Lemma \ref{lemma:NI property preservation}, then the nominal closed-loop system can be rendered to be an NI system whose transfer function is $R(s)=T_y^{-1}\hat R(s)T_y^{-T}$. Since $\mc Y_2$ and $\mc Y_3$ can be any positive definite matrices, we choose them to be such that
\begin{equation}\label{eq:DC gain condition transformed}
\lambda_{max}\left(T_y^{-1} \left[\begin{matrix}
\mc Y_2 & 0 \\ 0& \mc Y_3	
\end{matrix}
\right] T_y^{-T}\right) < \frac{1}{\gamma}.	
\end{equation}
Therefore, $\lambda_{max}(R(0)) < \frac{1}{\gamma}$. Hence, $\lambda_{max}( R(0)\Delta(0)) < 1$. According to Lemma \ref{lemma:dc gain theorem}, it now follows that the system (\ref{eq:original uncertain system}) is asymptotically stable. This completes the proof.
\end{IEEEproof}

\begin{remark}
In the case that the uncertainty (\ref{eq:uncertain SNI}) in the system (\ref{eq:original uncertain system}) is NI, we can render the nominal closed-loop system (\ref{eq:uncertain A}) and (\ref{eq:uncertain B}) OSNI using the results of Theorem \ref{thm:feedback OSNI} in order to achieve stabilization (see \cite{bhowmick2017lti,bhowmick2019output} for the corresponding OSNI stability results).
\end{remark}

\section{ILLUSTRATIVE EXAMPLE}\label{section:example}
In this section, we demonstrate the procedure of stabilizing an uncertain system by rendering the nominal closed-loop system NI with the DC gain conditions of Lemma \ref{lemma:dc gain theorem} satisfied. Consider an uncertain system with the following state-space model:
\begin{subequations}\label{eq:uncertain eg}
\begin{align}
	\dot x =&\ \left[\begin{matrix}
-1 & 0 & 1 & 1\\ 1 & -1 & 0 & 1\\ 1 & -1 & 1 & 0\\ 0 & 1 & -1 & 1	
\end{matrix}
\right]x+\left[\begin{matrix}
0 & 0 \\ 1 & 0\\ 1 & 0 \\ 1 & 1
\end{matrix}
\right](w + u),\label{eq:uncertain eg state}\\
	y =&\ \left[\begin{matrix}
0 & 1 & 0 & 0\\ 0 & 0 & 1 & 0
\end{matrix}
\right] x,\label{eq:uncertain eg output}\\
	w =&\ \Delta y,\label{eq:uncertain eg uncertainty}
\end{align}	
\end{subequations}
where $x\in \mathbb R^4$, $u\in \mathbb R^{2}$ and $y\in \mathbb R^{2}$ are the state, input and output of the nominal plant. Here, (\ref{eq:uncertain eg uncertainty}) models the uncertainty, and the uncertainty transfer function $\Delta(s)$ is assumed to be SNI with $\Delta(\infty)\geq 0$ and $\lambda_{max}(\Delta(0))\leq 1$. We aim to find a state feedback control law such that the system (\ref{eq:uncertain eg}) is asymptotically stable. Let us define the following:
\begin{align*}
\mc A=&	\left[\begin{matrix}
-1 & 0 & 1 & 1\\ 1 & -1 & 0 & 1\\ 1 & -1 & 1 & 0\\ 0 & 1 & -1 & 1	
\end{matrix}
\right],\\
\mc B=&\left[\begin{matrix}
0 & 0 \\ 1 & 0\\ 1 & 0 \\ 1 & 1
\end{matrix}
\right],\\
\mc C=& \left[\begin{matrix}
0 & 1 & 0 & 0\\ 0 & 0 & 1 & 0
\end{matrix}
\right].
\end{align*}
We have that $\mc C \mc B = \left[\begin{matrix}
 1 & 0\\ 1 & 0
\end{matrix}
\right]$, which is singular. However, the system can be output transformed into a form with a relative degree vector. We use the output transformation $\tilde y = T_y y$ with $T_y = \left[\begin{matrix}
 1 & 0\\ -1 & 1
\end{matrix}
\right]$. The transformed system has a relative degree vector $r=\{1,2\}$. To transform the system into its normal form as shown in (\ref{eq:original normal}), we also use a state transformation $\tilde x = T_x x$ with $T_x=\left[\begin{matrix}
 1 & 0 & 0 & 0\\ 0 & 1 & 0 & 0\\0&-1&1&0\\0&0&1&-1
\end{matrix}
\right]$ and an input transformation $\tilde u = T_u u$ with $T_u = \left[\begin{matrix}
 1 & 0\\ 0 & -1
\end{matrix}
\right]$. Letting $w=0$, the transformed system has the state-space realization:
\begin{subequations}\label{eq:eg normal}
\begin{align}
\tilde x =& \left[\begin{matrix}
 -1 & 2 & 2 & -1\\ 1 & 0 & 1 & -1\\0&0&0&1\\1&-1&1&1
\end{matrix}
\right]\tilde x + \left[\begin{matrix}
0 & 0 \\ 1 & 0\\ 0 & 0 \\ 0 & 1
\end{matrix}
\right]\tilde u,\\
\tilde y =& \left[\begin{matrix}
0 & 1 & 0 & 0\\ 0 & 0 & 1 & 0
\end{matrix}
\right]\tilde x,
\end{align}
\end{subequations}
where $\tilde x = \left[\begin{matrix}
z \\ \tilde x_1 \\ \tilde x_2 \\ \tilde x_3
\end{matrix}
\right]$ is the state, $\tilde u = \left[\begin{matrix}
\tilde u_1 \\ \tilde u_2
\end{matrix}
\right]$ is the input and $\tilde y = \left[\begin{matrix}
\tilde y_1 \\ \tilde y_2
\end{matrix}
\right]$ is the output. The system (\ref{eq:eg normal}) is in a normal form and it can be verified that it is minimal. It can be also observed that it has no zero at the origin and is weakly minimum phase. Therefore, according to Theorem \ref{thm:stabilization}, the uncertain system (\ref{eq:original uncertain system}) is stabilizable. We construct the state feedback control law according to the proof in Theorem \ref{thm:feedback NI}. Comparing the normal form (\ref{eq:eg normal}) of the example to the general normal form (\ref{eq:original normal}), we have that $A_{00}=-1$, $A_{01}=2$, $A_{02}=2$ and $A_{03}=-1$ in the system (\ref{eq:eg normal}). Then, using the formulas (\ref{eq:rd1 K_20})-(\ref{eq:K20b}), (\ref{eq:rd1 K_10})-(\ref{eq:K10b}) and (\ref{eq:K_11})-(\ref{eq:K_23}) with $\mc Y_1^b = 1$, $\mc H=1$ and $K_{13}=1$, we obtain that $K_{10}=1$, $K_{11}=-2-\frac{1}{\mc Y_2}$, $K_{12}=-2$, $K_{20}=4$, $K_{21}=-8$, $K_{22}=-8-\frac{1}{\mc Y_3}$ and $K_{23}=-\frac{1}{2}$. Then, choose the control inputs $\tilde u_1$ and $\tilde u_2$ as given in (\ref{eq:u1}) and (\ref{eq:u2}). That is
\begin{align*}
\tilde u_1 =&\ \tilde v_1+(K_{10}-1)z+K_{11}\tilde x_1+(K_{12}-1)\tilde x_2\\
&+(K_{13}+1)\tilde x_3,\notag\\
=&\  \tilde v_1-\left(2+\frac{1}{\mc Y_2}\right)\tilde x_1-3\tilde x_2+2\tilde x_3,
\end{align*}
where $\mc Y_2>0$ is a scalar, and
\begin{align*}
\tilde u_2=&\ \tilde v_2+(K_{20}-1)z+(K_{21}+1)\tilde x_1+(K_{22}-1)\tilde x_2\\
&+(K_{23}-1)\tilde x_3,\notag\\
=&\ \tilde v_2+3z-7\tilde x_1-\left(9+\frac{1}{\mc Y_3}\right)\tilde x_2-\frac{3}{2}\tilde x_3,
\end{align*}
where $\mc Y_3>0$ is a scalar. Here, $\tilde v = \left[\begin{matrix}
\tilde v_1 \\ \tilde v_2	
\end{matrix}
\right]$ is the new input of the output transformed system such that the transfer function from $\tilde v$ to $\tilde y$ is NI. To ensure that the DC gain of the closed-loop system (\ref{eq:uncertain eg}) is less than unity, we need to satisfy (\ref{eq:DC gain condition transformed}). A suitable choice is $\mc Y_2 = \frac{1}{4}$ and $\mc Y_3 = \frac{1}{4}$. Therefore, we have that
\begin{equation*}
\tilde u = \left[\begin{matrix}
0 & -6 & -3 & 2\\
3 & -7 & -13 & -\frac{3}{2}	
\end{matrix}
\right]	\tilde x + \tilde v.
\end{equation*}
According to Lemma \ref{lem:feedback NI irrelevance}, the NI property of the system from input $\tilde v$ to output $\tilde y$ implies that the system from input $v = T_y^T\tilde v$ to $y$ is also NI. Therefore, we choose the control input of the nominal plant (\ref{eq:uncertain eg state}) and (\ref{eq:uncertain eg output}) to be
\begin{align*}
u =&\  T_u^{-1} \tilde u \notag\\
 =&\ T_u^{-1}	\left[\begin{matrix}
0 & -6 & -3 & 2\\
3 & -7 & -13 & -\frac{3}{2}	
\end{matrix}
\right]	\tilde x + T_u^{-1} \tilde v \notag\\
 =&\ T_u^{-1}	\left[\begin{matrix}
0 & -6 & -3 & 2\\
3 & -7 & -13 & -\frac{3}{2}	
\end{matrix}
\right] T_x x + T_u^{-1} T_y^{-T} v\notag\\
=&\ \left[\begin{matrix}
0 & -3 & -1 & -2\\
-3 & -6 & 14.5 & -1.5	
\end{matrix}
\right]x+\left[\begin{matrix}
1&1\\0&-1
\end{matrix}
\right]v.
\end{align*}
Under this input, the nominal plant given in (\ref{eq:uncertain eg state}) and (\ref{eq:uncertain eg output}) with $w=0$ is NI. In the case that $w$ is regarded as the external input of the nominal plant (\ref{eq:uncertain eg state}) and (\ref{eq:uncertain eg output}), following the proof of Theorem \ref{thm:stabilization}, let
\begin{align*}
u=& \left[\begin{matrix}
0 & -3 & -1 & -2\\
-3 & -6 & 14.5 & -1.5	
\end{matrix}
\right]x+\left(\left[\begin{matrix}
1&1\\0&-1
\end{matrix}
\right]-I\right)w\notag\\
=&\left[\begin{matrix}
0 & -3 & -1 & -2\\
-3 & -6 & 14.5 & -1.5
\end{matrix}
\right]x+\left[\begin{matrix}
0&1\\0&-2
\end{matrix}
\right]w.
\end{align*}
With this control input, the uncertain system (\ref{eq:uncertain eg}) becomes
\begin{subequations}\label{eq:eg controlled}
\begin{align}
	\dot x =&\ \left[\begin{matrix}
-1 & 0 & 1 &1\\ 1 & -4 & -1 & -1\\ 1 & -4 & 0 & -2\\ -3 & -8 & 12.5 & -2.5	
\end{matrix}
\right]x+\left[\begin{matrix}
0 & 0 \\ 1 & 1\\ 1 & 1 \\ 1 & 0
\end{matrix}
\right]w,\label{eq:eg controlled state}\\
	y =&\ \left[\begin{matrix}
0 & 1 & 0 & 0\\ 0 & 0 & 1 & 0
\end{matrix}
\right] x,\label{eq:eg controlled output}\\
	w =&\ \Delta y.\label{eq:eg controlled uncertainty}
\end{align}	
\end{subequations}
The transfer function matrix of the nominal closed-loop system (\ref{eq:eg controlled state}) and (\ref{eq:eg controlled output}) is
\begin{align*}
R(s)=&\	\frac{1}{2s^4+15s^3+63s^2+156s+32}\times \\
&\left[\begin{matrix}
2s^3+3s^2+33s+8 & 2s^3+5s^2+29s+8 \\
2s^3+3s^2+17s+8 & 2s^3+7s^2+27s+16
\end{matrix}
\right].
\end{align*}
It can be verified that all poles of this transfer function matrix have negative real parts. Also, we have $j[R(j\omega)-R(j\omega)^*]\geq 0$ for all $\omega>0$. Therefore, the nominal closed-loop system (\ref{eq:eg controlled state}) and (\ref{eq:eg controlled output}) is NI. In addition, we have that
\begin{equation*}
R(0)=	\left[\begin{matrix}
0.25 & 0.25 \\ 0.25 & 0.5
\end{matrix}
\right].
\end{equation*}
Hence, $\lambda_{max}(R(0))=0.6545$. Therefore, $\lambda_{max}(R(0)\Delta(0))\leq \lambda_{max}(R(0)) \lambda_{max}(\Delta(0))<1$. Hence, according to Lemma \ref{lemma:dc gain theorem}, the system (\ref{eq:eg controlled}) is asymptotically stable.

\section{CONCLUSION}
\label{section:conclusion}
In this paper, we have provided necessary and sufficient conditions under which a linear system can be rendered NI. As stated in Theorem \ref{thm:feedback NI}, a minimal linear system (\ref{eq:original}) with no zeros at the origin is state feedback equivalent to an NI system if and only if it can be output transformed to a system, which has relative degree less than or equal to two and is weakly minimum phase. Similar OSNI and SSNI state feedback equivalence results are presented in Theorems \ref{thm:feedback OSNI} and \ref{thm:feedback SSNI}. The NI state feedback equivalence results are then applied to robustly stabilize a system with SNI uncertainty. An example is also provided to illustrate the process of rendering a system NI in order to stabilize an uncertain system.

\bibliographystyle{IEEEtran}

\begin{thebibliography}{10}
\providecommand{\url}[1]{#1}
\csname url@samestyle\endcsname
\providecommand{\newblock}{\relax}
\providecommand{\bibinfo}[2]{#2}
\providecommand{\BIBentrySTDinterwordspacing}{\spaceskip=0pt\relax}
\providecommand{\BIBentryALTinterwordstretchfactor}{4}
\providecommand{\BIBentryALTinterwordspacing}{\spaceskip=\fontdimen2\font plus
\BIBentryALTinterwordstretchfactor\fontdimen3\font minus
  \fontdimen4\font\relax}
\providecommand{\BIBforeignlanguage}[2]{{%
\expandafter\ifx\csname l@#1\endcsname\relax
\typeout{** WARNING: IEEEtran.bst: No hyphenation pattern has been}%
\typeout{** loaded for the language `#1'. Using the pattern for}%
\typeout{** the default language instead.}%
\else
\language=\csname l@#1\endcsname
\fi
#2}}
\providecommand{\BIBdecl}{\relax}
\BIBdecl

\bibitem{lanzon2008stability}
A.~Lanzon and I.~R. Petersen, ``Stability robustness of a feedback
  interconnection of systems with negative imaginary frequency response,''
  \emph{IEEE Transactions on Automatic Control}, vol.~53, no.~4, pp.
  1042--1046, 2008.

\bibitem{petersen2010feedback}
I.~R. Petersen and A.~Lanzon, ``Feedback control of negative-imaginary
  systems,'' \emph{IEEE Control Systems Magazine}, vol.~30, no.~5, pp. 54--72,
  2010.

\bibitem{xiong2010negative}
J.~Xiong, I.~R. Petersen, and A.~Lanzon, ``A negative imaginary lemma and the
  stability of interconnections of linear negative imaginary systems,''
  \emph{IEEE Transactions on Automatic Control}, vol.~55, no.~10, pp.
  2342--2347, 2010.

\bibitem{song2012negative}
Z.~Song, A.~Lanzon, S.~Patra, and I.~R. Petersen, ``A negative-imaginary lemma
  without minimality assumptions and robust state-feedback synthesis for
  uncertain negative-imaginary systems,'' \emph{Systems \& Control Letters},
  vol.~61, no.~12, pp. 1269--1276, 2012.

\bibitem{mabrok2014generalizing}
M.~A. Mabrok, A.~G. Kallapur, I.~R. Petersen, and A.~Lanzon, ``Generalizing
  negative imaginary systems theory to include free body dynamics: Control of
  highly resonant structures with free body motion,'' \emph{IEEE Transactions
  on Automatic Control}, vol.~59, no.~10, pp. 2692--2707, 2014.

\bibitem{wang2015robust}
J.~Wang, A.~Lanzon, and I.~R. Petersen, ``Robust cooperative control of
  multiple heterogeneous negative-imaginary systems,'' \emph{Automatica},
  vol.~61, pp. 64--72, 2015.

\bibitem{bhikkaji2011negative}
B.~Bhikkaji, S.~R. Moheimani, and I.~R. Petersen, ``A negative imaginary
  approach to modeling and control of a collocated structure,'' \emph{IEEE/ASME
  Transactions on Mechatronics}, vol.~17, no.~4, pp. 717--727, 2011.

\bibitem{bhowmick2017lti}
P.~Bhowmick and S.~Patra, ``On {LTI} output strictly negative-imaginary
  systems,'' \emph{Systems \& Control Letters}, vol. 100, pp. 32--42, 2017.

\bibitem{preumont2018vibration}
A.~Preumont, \emph{Vibration control of active structures: an
  introduction}.\hskip 1em plus 0.5em minus 0.4em\relax Springer, 2018, vol.
  246.

\bibitem{halim2001spatial}
D.~Halim and S.~R. Moheimani, ``Spatial resonant control of flexible
  structures-application to a piezoelectric laminate beam,'' \emph{IEEE
  Transactions on Control Systems Technology}, vol.~9, no.~1, pp. 37--53, 2001.

\bibitem{pota2002resonant}
H.~Pota, S.~R. Moheimani, and M.~Smith, ``Resonant controllers for smart
  structures,'' \emph{Smart Materials and Structures}, vol.~11, no.~1, p.~1,
  2002.

\bibitem{mabrok2013spectral}
M.~A. Mabrok, A.~G. Kallapur, I.~R. Petersen, and A.~Lanzon, ``Spectral
  conditions for negative imaginary systems with applications to
  nanopositioning,'' \emph{IEEE/ASME Transactions on Mechatronics}, vol.~19,
  no.~3, pp. 895--903, 2013.

\bibitem{das2014mimo}
S.~K. Das, H.~R. Pota, and I.~R. Petersen, ``A {MIMO} double resonant
  controller design for nanopositioners,'' \emph{IEEE Transactions on
  Nanotechnology}, vol.~14, no.~2, pp. 224--237, 2014.

\bibitem{das2014resonant}
------, ``Resonant controller design for a piezoelectric tube scanner: A mixed
  negative-imaginary and small-gain approach,'' \emph{IEEE Transactions on
  Control Systems Technology}, vol.~22, no.~5, pp. 1899--1906, 2014.

\bibitem{das2015multivariable}
------, ``Multivariable negative-imaginary controller design for damping and
  cross coupling reduction of nanopositioners: a reference model matching
  approach,'' \emph{IEEE/ASME Transactions on Mechatronics}, vol.~20, no.~6,
  pp. 3123--3134, 2015.

\bibitem{cai2010stability}
C.~Cai and G.~Hagen, ``Stability analysis for a string of coupled stable
  subsystems with negative imaginary frequency response,'' \emph{IEEE
  Transactions on Automatic Control}, vol.~55, no.~8, pp. 1958--1963, 2010.

\bibitem{rahman2015design}
M.~A. Rahman, A.~Al~Mamun, K.~Yao, and S.~K. Das, ``Design and implementation
  of feedback resonance compensator in hard disk drive servo system: A mixed
  passivity, negative-imaginary and small-gain approach in discrete time,''
  \emph{Journal of Control, Automation and Electrical Systems}, vol.~26, no.~4,
  pp. 390--402, 2015.

\bibitem{brogliato2007dissipative}
B.~Brogliato, R.~Lozano, B.~Maschke, and O.~Egeland, ``Dissipative systems
  analysis and control,'' \emph{Theory and Applications}, vol.~2, 2007.

\bibitem{mabrok2020dissipativity}
M.~A. Mabrok, M.~A. Alyami, and E.~E. Mahmoud, ``On the dissipativity property
  of negative imaginary systems,'' \emph{Alexandria Engineering Journal}, 2020.

\bibitem{kokotovic1989positive}
P.~Kokotovic and H.~Sussmann, ``A positive real condition for global
  stabilization of nonlinear systems,'' \emph{Systems \& Control Letters},
  vol.~13, no.~2, pp. 125--133, 1989.

\bibitem{saberi1990global}
A.~Saberi, P.~Kokotovic, and H.~Sussmann, ``Global stabilization of partially
  linear composite systems,'' \emph{SIAM Journal on Control and Optimization},
  vol.~28, no.~6, pp. 1491--1503, 1990.

\bibitem{byrnes1991passivity}
C.~Byrnes, A.~Isidori, and J.~Willems, ``Passivity, feedback equivalence, and
  the global stabilization of minimum phase nonlinear systems,'' \emph{IEEE
  Transactions on Automatic Control}, vol.~36, no.~11, pp. 1228--1240, 1991.

\bibitem{byrnes1991asymptotic}
C.~I. Byrnes and A.~Isidori, ``Asymptotic stabilization of minimum phase
  nonlinear systems,'' \emph{IEEE Transactions on Automatic Control}, vol.~36,
  no.~10, pp. 1122--1137, 1991.

\bibitem{santosuosso1997passivity}
G.~Santosuosso, ``Passivity of nonlinear systems with input-output
  feedthrough,'' \emph{Automatica}, vol.~33, no.~4, pp. 693--697, 1997.

\bibitem{lin1995feedback}
W.~Lin, ``Feedback stabilization of general nonlinear control systems: a
  passive system approach,'' \emph{Systems \& Control Letters}, vol.~25, no.~1,
  pp. 41--52, 1995.

\bibitem{jiang1996passification}
Z.-P. Jiang, D.~J. Hill, and A.~L. Fradkov, ``A passification approach to
  adaptive nonlinear stabilization,'' \emph{Systems \& Control Letters},
  vol.~28, no.~2, pp. 73--84, 1996.

\bibitem{isidori2013nonlinear}
A.~Isidori, \emph{Nonlinear control systems}.\hskip 1em plus 0.5em minus
  0.4em\relax Springer Science \& Business Media, 2013.

\bibitem{shi2021negative}
K.~Shi, I.~R. Petersen, and I.~G. Vladimirov, ``Negative imaginary state
  feedback equivalence for systems of relative degree one and relative degree
  two,'' \emph{To appear in the proceedings of the 60th IEEE Conference on
  Decision and Control (CDC 2021), Dec 2021, Austin, Texas, USA, Full version
  available as arXiv preprint arXiv:2103.05249}, 2021.

\bibitem{bhowmick2019output}
P.~Bhowmick and A.~Lanzon, ``Output strictly negative imaginary systems and its
  connections to dissipativity theory,'' in \emph{2019 IEEE 58th Conference on
  Decision and Control (CDC)}.\hskip 1em plus 0.5em minus 0.4em\relax IEEE,
  2019, pp. 6754--6759.

\bibitem{lanzon2011strongly}
A.~Lanzon, S.~Patra, I.~R. Petersen, and Z.~Song, ``A strongly strict
  negative-imaginary lemma for non-minimal linear systems,''
  \emph{Communications in Information and Systems}, vol.~11, no.~2, pp.
  139--142, 2011.

\bibitem{bernstein2009matrix}
D.~S. Bernstein, \emph{Matrix mathematics: theory, facts, and formulas}.\hskip
  1em plus 0.5em minus 0.4em\relax Princeton university press, 2009.

\bibitem{hespanha2018linear}
J.~P. Hespanha, \emph{Linear systems theory}.\hskip 1em plus 0.5em minus
  0.4em\relax Princeton university press, 2018.

\bibitem{fomichev2016generalization}
V.~Fomichev, A.~Kraev, and A.~Rogovskii, ``Generalization of the notion of
  relative degree and its properties,'' \emph{Differential Equations}, vol.~52,
  no.~8, pp. 1061--1071, 2016.

\bibitem{kraev2014generalization}
A.~Kraev, A.~Rogovskii, and V.~Fomichev, ``On a generalization of relative
  degree,'' \emph{Differential Equations}, vol.~50, no.~8, pp. 1122--1127,
  2014.

\bibitem{khalil2002nonlinear}
H.~K. Khalil and J.~W. Grizzle, \emph{Nonlinear systems}.\hskip 1em plus 0.5em
  minus 0.4em\relax Prentice hall Upper Saddle River, NJ, 2002, vol.~3.

\end{thebibliography}

\end{document}